\documentclass[11pt,a4paper]{article}
\pdfoutput=1
\usepackage{jcappub}
\usepackage{rotating}
\usepackage{array}
\usepackage{amsmath}
\usepackage[normalem]{ulem}
\usepackage{slashed}
\usepackage{booktabs}
\usepackage[pdftex,table]{xcolor}
\usepackage{units}
\usepackage{xspace}
\usepackage{makecell}

\newcommand{\ba}[1]{\ensuremath{\left( #1 \right)}}
\newcommand{\bb}[1]{\ensuremath{\left[ #1 \right]}}

\newcommand{\pd}[2]{\ensuremath{\frac{\partial #1}{\partial #2}}}

\newcommand{\td}[2]{\ensuremath{\frac{\mathrm d #1}{\mathrm d #2}}}
\newcommand{\tdd}[2]{\ensuremath{\frac{\mathrm d^2 #1}{\mathrm d #2^2}}} 

\newcommand{\nocontentsline}[3]{}
\newcommand{\tocless}[2]{\bgroup\let\addcontentsline=\nocontentsline#1{#2}\egroup}

\newcommand*\diff{\mathop{}\!\mathrm{d}}

\newcommand{\Mp}{\ensuremath{m_\text{Pl}}}

\newcommand{\grhoSM}{\ensuremath{ {g_{\text{eff},\rho}^\text{SM}}}\xspace}
\newcommand{\gsSM}{\ensuremath{{g_{\text{eff},s}^\text{SM}}}\xspace}
\newcommand{\grhoSMcd}{\ensuremath{ {g_{\text{eff},\rho}^\text{SM,cd}}}\xspace}
\newcommand{\gsSMcd}{\ensuremath{{g_{\text{eff},s}^\text{SM,cd}}}\xspace}

\usepackage{multirow}

\arxivnumber{TTK-21-36, DESY-22-014}

\title{Turn up the volume: listening to phase transitions in hot dark sectors}

\author[1]{Fatih Ertas,}
\author[1]{Felix Kahlhoefer}
\author[1,2]{and Carlo Tasillo}

\affiliation[1]{Institute for Theoretical Particle Physics and Cosmology (TTK),\\RWTH Aachen University,\\D-52056 Aachen, Germany}

\affiliation[2]{Deutsches Elektronen-Synchrotron DESY,\\ Notkestr. 85, 22607 Hamburg, Germany}

\emailAdd{ertas@physik.rwth-aachen.de}
\emailAdd{kahlhoefer@physik.rwth-aachen.de}
\emailAdd{carlo.tasillo@desy.de}

\abstract{Stochastic gravitational wave (GW) backgrounds from first-order phase transitions are an exciting target for future GW observatories and may enable us to study dark sectors with very weak couplings to the Standard Model. In this work we show that such signals may be significantly enhanced for hot dark sectors with a temperature larger than the one of the SM thermal bath. The need to transfer the entropy from the dark sector to the SM after the phase transition can however lead to a substantial dilution of the GW signal. We study this dilution in detail, including the effect of number-changing processes in the dark sector (so-called cannibalism), and show that in large regions of parameter space a net enhancement remains. We apply our findings to a specific example of a dark sector containing a dark Higgs boson and a dark photon and find excellent detection prospects for LISA and the Einstein telescope.}


\begin{document}
\maketitle
\flushbottom

\section{Introduction}
\label{sec:introduction}

The observation of gravitational wave (GW) signals from binary mergers has opened up a new window to the universe~\cite{LIGOScientific:2016aoc,LIGOScientific:2018mvr,Abbott2020}. Not only can we expect major breakthroughs in our understanding of compact astrophysical objects, but future GW observatories will enable us to explore the early universe beyond the Cosmic Microwave Background. One of the most exciting prospects is the detection of a stochastic GW background of cosmological origin, which may arise for example from inflation, cosmic strings or a first-order phase transition~\cite{Maggiore:2018sht,Bertone:2019irm}.

A particularly attractive possibility is that future GW observatories may search for GWs produced by phase transitions within a dark sector~\cite{Schwaller:2015tja,Caprini2015,Caprini2019,Huang:2021rrk}. The presence of such dark sectors is well-motivated by the need to explain the role of dark matter in the early universe and during structure formation. Since any attempts to discover dark matter in the laboratory have so far been unsuccessful, it is a plausible possibility that the dark sector only interacts very feebly with Standard Model particles. In such a case, gravitational wave signals may offer unique opportunities to study the structure and dynamics of dark sectors.

If the dark sector is not in thermal contact with the Standard Model, its temperature $T_\mathrm{DS}$ may be different from the temperature of the thermal bath of SM particles $T_\mathrm{SM}$. Indeed, the larger the ratio $\xi = T_\mathrm{DS}/T_\mathrm{SM}$ the more energy is stored in the dark sector and can be released during a phase transition in the form of gravitational wave signals~\cite{Breitbach2019}. In the present work we therefore focus on GW signals from \emph{hot} dark sectors with $\xi > 1$. Such a temperature difference could be a direct result of the details of reheating~\cite{Garani:2021zrr}, but it could also be generated much later, for example if heavy particles in the dark sector annihilate and transfer their entropy to the lighter degrees of freedom. 

However, an often overlooked problem is what happens to the energy density of a decoupled dark sector after the end of the phase transition. If the dark sector contains any light or massless states, measurements of the number of relativistic degrees of freedom during Big Bang Nucleosynthesis (BBN) and recombination place strong bounds on the temperature ratio $\xi$ (see e.g.~\cite{GAMBITCosmologyWorkgroup:2020htv}). In the presence of massive stable states, on the other hand, the universe would typically enter matter domination much earlier than observed. Based on this line of reasoning, it was argued in Ref.~\cite{Breitbach2019} that the dark sector should be much colder than the SM, which in turn places a strong bound on the magnitude of any gravitational wave signal (see also Ref.~\cite{Fairbairn:2019xog}). 

Here we consider an alternative possibility, namely that the energy of the dark sector is transferred to the SM via out-of-equilibrium decays of the lightest dark sector particle. Such decays inject entropy into the SM thermal bath and thereby alter the expansion history of the universe. Indeed, such decays have been studied as a possibility of decreasing the dark matter relic abundance after these particles have decoupled from the thermal bath \cite{Scherrer1985,McDonald:1989jd,Giudice:2000ex,Berlin:2016gtr,Bramante:2017obj,Cirelli2018,Evans:2019jcs,Heurtier:2019eou}. In a similar fashion, entropy injection leads to additional red-shifting, i.\,e.\ an effective dilution, of stochastic GW backgrounds~\cite{Ellis:2020nnr}.

To investigate these effects in detail, we perform a model-independent calculation of the effect of out-of-equilibrium decays on GW signals in terms of the properties and the abundance of the decaying particles. We improve upon previous studies by including the effect of number-changing processes (so-called cannibalism~\cite{Carlson:1992fn,Pappadopulo:2016pkp,Farina:2016llk,Buen-Abad:2018mas,Erickcek:2020wzd,Heimersheim:2020aoc}) when the lightest dark sector particle becomes non-relativistic. We then apply our results to a specific dark sector model of a dark photon coupled to a dark Higgs field, which develops a non-zero vacuum expectation value (vev) through a first-order phase transition. In the parameter region where the phase transition is strongest, the lightest dark sector particle is the dark Higgs boson, which can then decay for example via a tiny mixing with the SM Higgs boson.

\begin{figure}
	\centering
	\includegraphics[width=0.8\linewidth]{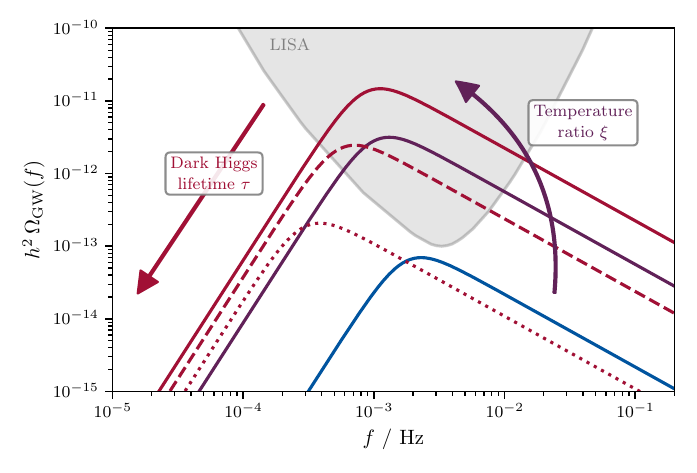}
	\caption{Example for the effects of the dark sector temperature ratio $\xi$ and the dark Higgs lifetime $\tau$ on the stochastic GW background spectrum. An increase in $\xi$ increases the transition strength and thereby amplifies the signal. A long-lived dark Higgs however injects a considerable amount of entropy into the SM bath, which dilutes the signal. The specific scenario considered here can be tested by LISA for sufficiently large temperature ratios and small lifetimes as indicated by the gray shaded power-law integrated sensitivity curve.}
	\label{fig:overviewplot1}
\end{figure}

We estimate the resulting GW signals and the corresponding signal-to-noise ratios and find that it is possible in this set-up to produce observable signals in planned GW observatories such as LISA~\cite{2017arXiv170200786A,Flauger:2020qyi} and the Einstein Telescope (ET)~\cite{Maggiore:2019uih}. These signals can be enhanced for temperature ratios $\xi > 1$ and an excessive dilution of the signal can be avoided if the dark Higgs boson decays sufficiently quickly after the phase transition. This is illustrated in figure~\ref{fig:overviewplot1}, which shows an example of a GW signal produced by a dark sector phase transition, as well as its dependence on the temperature ratio $\xi$ and the lifetime $\tau$ of the dark Higgs boson. In this example, if $\xi$ is sufficiently large and $\tau$ is sufficiently small, the GW signal can be substantially enhanced and can lie within the projected sensitivity of LISA. These conclusions are very general and apply to other types of dark sectors as well as more refined calculations of GW signals from first-order phase transitions.

The remainder of this work is structured as follows. In section~\ref{sec:darksector} we introduce the general formalism for describing a hot dark sector that is not in thermal equilibrium with the SM thermal bath. We also review the calculation of the effective potential and of the stochastic GW background arising from strong first-order phase transitions. In section~\ref{sec:dilution} we then consider the subsequent evolution of the dark sector and how the transfer of entropy to the SM bath leads to a dilution of GW signals. Finally, we apply this general discussion to a specific dark sector model in section~\ref{sec:model} and obtain the predicted signal-to-noise ratios in future GW observatories. The code used to obtain our results, described in detail in appendix~\ref{app:updates}, is publicly available as \textsc{TransitionListener} at \url{https://github.com/tasicarl/TransitionListener}.

\section{General formalism}
\label{sec:darksector}

We begin this section by reviewing some relevant concepts from cosmology and introducing our notation for describing dark sectors. We then briefly present the calculation of the effective potential for a dark Higgs field at finite temperatures and how this can give rise to a first-order phase transition. Finally, we discuss the resulting stochastic GW background and summarize the approximations made in the present work.

\subsection{Dark sector cosmology}

In a flat Friedmann-Lema\^{i}tre-Robertson-Walker universe, the Friedmann equations read
\begin{align}
H(t) \equiv \frac{\dot{a}(t)}{a(t)} = \sqrt{\frac{\rho_\text{tot}(t)}{3 \, \Mp^2}} \; , & &
\dot{\rho}_\text{tot}(t) + 3 \, H(t) \,  \bb{\rho_\text{tot}(t) + P_\text{tot}(t)} = 0 \; . \label{eq:Friedmann12}
\end{align}
Here and in the following, $m_\text{Pl} = \ba{8 \, \pi \, G}^{-1/2} \simeq 2 \cdot 10^{18} \, \text{GeV}$ denotes the reduced Planck mass. The Hubble rate $H(t)$ works as a measure for the expansion rate of the universe and can be calculated using the total energy density $\rho_\text{tot}(t)$ of the primordial plasma. The time evolution of $\rho_\text{tot}(t)$ in an expanding universe is described by the second Friedmann equation, where $P_\text{tot}(t)$ denotes the pressure of the primordial plasma. The total energy density and pressure can be obtained by summing over the contributions from all individual particle species $x$, that is $\rho_\text{tot}(t) = \sum_x \rho_x(t)$ and $P_\text{tot}(t) = \sum_x P_x(t)$. We further introduce the volume heating rate \cite{Hufnagel2020}
\begin{align}
\label{eq:vol_heat}
\dot{q}_x(t) = \dot{\rho}_x(t) + 3 \, H(t) \, [\rho_x(t) + P_x(t)] \; ,
\end{align}
for a given particle species $x$. The second Friedmann equation therefore states that the total heat is conserved: $\dot{Q}_\text{tot} = \sum_x \dot{Q}_x = 0$, where $\dot{Q}_x = \dot{q}_x \, a^3(t)$.

To make use of the Friedmann equations, the time dependences of $\rho_\text{tot}(t)$ and $P_\text{tot}(t)$, and therefore of all individual $\rho_x(t)$ and $P_x(t)$, have to be known. The full evolution of these thermodynamical quantities for a given particle species $x$ is encoded in its distribution function $f_x(t,p)$. If particles of the species $x$ scatter frequently enough, they will thermalize and $f_x(t,p)$ will follow a Bose-Einstein or Fermi-Dirac distribution. The thermal bath is then completely determined by its temperature $T_x(t)$ and the chemical potential $\mu_x(t)$. In the case of this so-called ``local thermal equilibrium'', the time-temperature relation $T_x(t)$ can be inverted and used to replace the time dependence in the previous functions by a temperature dependence.

The distribution function $f_x(t, p)$ can further be used to obtain the comoving entropy density $S_x(t)$ of a generic particle species $x$. If $x$ follows a local thermal equilibrium, one finds that the second law of thermodynamics holds individually for $x$, that is \cite{Kolb1990}
\begin{align}
T_x(t) \, \dot{S}_x(t) = \dot{Q}_x(t) - \mu_x(t) \, \dot{N}_x(t) \; , \label{eq:2ndlaw}
\end{align}
where $N_x(t) = n_x(t) \, a^3(t)$ is the comoving particle number density of $x$. Hence, the comoving entropy density $S_x$ is conserved, if $\dot{Q}_x = 0$ and $\mu_x(t) \, \dot{N}_x(t) = 0$. This is the case, when no heat is transferred between $x$ and other particle species and when either $\mu_x(t) = 0$ or $\dot{N}_x(t) = \dot{n}_x + 3 \, H(t) \, n_x(t) = 0$. Moreover, eq.~\eqref{eq:2ndlaw} can be used to define the entropy density $s_x(t) = S_x(t) / a^3(t)$, which implies \cite{Kolb1990}
\begin{align}
T_x(t) \, s_x(t) = \rho_x(t) + P_x(t) - \mu_x(t) \, n_x(t)\;,
\end{align}
in local thermal equilibrium.

If $\mu_x \ll T_x$, the thermal distribution functions $f_x(t,p)$ can therefore be integrated to obtain
\begin{subequations}
\label{eq:thermalizedquantities}
\begin{align}
\rho_x(T_x) &= \frac{g_x \, T_x^4}{2 \, \pi^2} \int_{z_x}^{\infty} \diff u_x \, \frac{u^2_x \sqrt{u^2_x - z_x^2}}{e^{u_x} \pm 1} \; ,\label{eq:thermalizedenergydensity}\\
P_x(T_x) &= \frac{g_x\, T_x^4}{6\, \pi^2} \int_{z_x}^{\infty} \diff u_x \, \frac{\ba{u^2_x - z_x^2}^{3/2}}{e^{u_x} \pm 1} \; ,\label{eq:thermalizedpressure}\\
s_x(T_x) &= \frac{g_x \,T_x^3}{2 \,\pi^2} \int_{z_x}^{\infty} \diff u_x \, \bb{\frac{u^2_x \sqrt{u^2_x - z_x^2}}{e^{u_x} \pm 1} + \frac{1}{3} \frac{\ba{u^2_x - z_x^2}^{3/2}}{e^{u_x} \pm 1}} \; . \label{eq:thermalizedentropydensity}
\end{align}
\end{subequations}
Here the substitutions $u_x = \sqrt{m_x^2 + p^2} / T_x$ and $z_x = m_x / T_x$ have been employed and a $+$ ($-$) sign refers to a fermionic (bosonic) species $x$. 

A handy feature of these equations is that one can introduce effective relativistic degrees of freedom, which can be used to elegantly express energy and entropy densities of thermal baths consisting of multiple particle species. Dividing $\rho_x(T_x)$ by $\left.\rho_\text{bos}^\text{rel}(T_x)\right|_{g=1} = \frac{\pi^2}{30} \, T_x^4$ and $P_x(T_x)$ by $\left.P_\text{bos}^\text{rel}(T_x)\right|_{g=1} = \frac{\pi^2}{90} \, T_x^4$, one can define \cite{Husdal2016}
\begin{subequations}
\begin{align}
g_{\text{eff},\rho}^x(T_x) &\equiv \frac{\rho_x(T_x)}{\left.\rho_\text{bos}^\text{rel}(T_x)\right|_{g=1}} = \frac{15 \,  g_x}{\pi^4}  \int_{z_x}^{\infty} \diff u_x \, \frac{u^2_x \sqrt{u^2_x - z_x^2}}{e^{u_x} \pm 1}\; ,\label{eq:geffrho}\\
g_{\text{eff},P}^x(T_x) &\equiv \frac{P_x(T_x)}{\left.P_\text{bos}^\text{rel}(T_x)\right|_{g=1}} = \frac{15 \, g_x}{\pi^4}  \int_{z_x}^{\infty} \diff u_x \, \frac{\ba{u^2_x - z_x^2}^{3/2}}{e^{u_x} \pm 1}\; ,\\
g_{\text{eff},s}^x(T_x) &= \frac{3 \, g_{\text{eff},\rho}^x(T_x) + g_{\text{eff},P}^x(T_x)}{4} \; . \label{eq:geffs}
\end{align}
\end{subequations}

\begin{figure}
\centering
\includegraphics[width=0.8\linewidth]{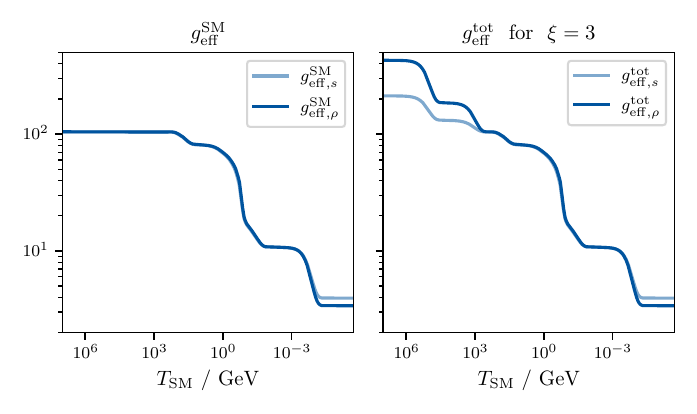}
\caption{The temperature evolution of the effective degrees of freedom of the SM (\textit{left}) and the total system, including also a dark sector (\textit{right}), assuming that the dark sector particle species follow their equilibrium distributions for all times. This dark sector consists of a dark photon with mass $m_{A^\prime} = 10^6\, \text{GeV}$ (and three internal degrees of freedom) and a dark Higgs with mass $m_\phi = 10^4\, \text{GeV}$ (and one internal degree of freedom). The temperature ratio between the two thermal baths was fixed to $\xi = 3$ to show that a dark sector slightly hotter than the SM bath can already yield interesting new dynamics. A possible temperature dependence of $\xi(T_\text{SM})$ as it would arise from the reheating of either sector was ignored here.}
\label{fig:geff}
\end{figure}

In the present work, we will consider a bath of SM particles and a separate thermal bath that we call the dark sector. As the interactions between the SM particles and the dark sector are assumed to be too feeble for the two sectors to thermalize, the two baths will in general have distinct temperatures $T_\text{SM}$ and $T_\text{DS}$ \cite{Breitbach2019}. Introducing the ratio $\xi = T_\text{DS} / T_\text{SM}$ of these temperatures, we find that the total energy and entropy densities of the primordial plasma is given by
\begin{subequations}
\begin{align}
\rho_\text{tot}(T_\text{SM}) &= \rho_\text{SM}(T_\text{SM}) + \rho_\text{DS}(T_\text{SM})
= \underbrace{\bb{g_{\text{eff},\rho}^\text{SM}(T_\text{SM}) + g_{\text{eff},\rho}^\text{DS}(T_\text{SM}) \, \xi^4(T_\text{SM}) }}_{\equiv g_{\text{eff},\rho}^\text{tot}(T_\text{SM})} \, \frac{\pi^2}{30} \, T_\text{SM}^4 \label{eq:rhotot} \; ,\\
s_\text{tot}(T_\text{SM}) &= s_\text{SM}(T_\text{SM}) + s_\text{DS}(T_\text{SM}) =  \underbrace{\bb{g_{\text{eff},s}^\text{SM}(T_\text{SM}) + g_{\text{eff},s}^\text{DS}(T_\text{SM}) \, \xi^3(T_\text{SM}) }}_{g_{\text{eff},s}^\text{tot}(T_\text{SM})} \, \frac{2\pi^2}{45} \, T_\text{SM}^3 \; , \label{eq:stot}
\end{align}
\end{subequations}
where $g_{\text{eff},\rho}^\text{SM}$ ($g_{\text{eff},s}^\text{SM}$) denotes the effective energy (entropy) degrees of freedom for the photon bath. The corresponding functions for the dark sector are denoted by the index ``DS''. Due to the high powers of $\xi$ that enter into eqs.~\eqref{eq:rhotot} and~\eqref{eq:stot}, the contributions from dark sectors that are only slightly hotter than the SM bath (i.\,e.\ $\xi > 1$) can have a large influence on $\rho_\text{tot}$ and $s_\text{tot}$. Hot dark sectors can therefore significantly modify the thermal history of the early universe. This effect is shown in figure~\ref{fig:geff} for two hot dark sector species (a dark Higgs boson and a dark photon, see section~\ref{sec:model}) in addition to the particles of the SM bath for $\xi = 3$. In this plot and the following work, we used the data for the SM effective degrees of freedom given in the ancillary material of Ref.~\cite{Saikawa2018}.

To describe the overall evolution of the combined system comprising the dark sector and the SM bath, we have to calculate the time (i.\,e.\ temperature) dependence of $\xi(T_\text{SM})$. For this purpose, we can use the fact that entropy is conserved individually in the two decoupled baths. Thus, $S_\text{SM} = \frac{2\pi^2}{45}\,  g_{\text{eff},s}^\text{SM} \, T_\text{SM}^3 \, a^3$ and  $S_\text{DS} = \frac{2\pi^2}{45}\,  g_{\text{eff},s}^\text{DS} \, T_\text{DS}^3 \, a^3$ are both constant and hence
\begin{align}
\xi(T_\text{SM}) = \xi(\tilde{T}_\text{SM}) \, \ba{\frac{g_{\text{eff},s}^\text{SM} (T_\text{SM})}{g_{\text{eff},s}^\text{SM}(\tilde{T}_\text{SM})}}^{1/3} \ba{\frac{g_{\text{eff},s}^\text{DS}(\tilde{T}_\text{SM})}{g_{\text{eff},s}^\text{DS} (T_\text{SM})}}^{1/3}.	\label{eq:selfconsistentxi}
\end{align}
Here, the quantity $\tilde{T}_\text{SM}$ specifies the temperature of the SM bath at a point in time where the two sectors have already been decoupled. Note that $g_{\text{eff},s}^\text{DS}(T_\text{SM})$ depends implicitly on $\xi(T_\text{SM})$ and therefore eq.~\eqref{eq:selfconsistentxi} must in general be solved numerically. The general result is that the temperature ratio $\xi$ increases when the dark sector degrees of freedom $g_{\text{eff},s}^\text{DS}$ decrease and that it decreases when the SM degrees of freedom $g_{\text{eff},s}^\text{SM}$ decrease.

To conclude this discussion we emphasize that while the description we have presented above is valid for large parts of the thermal history, we will also encounter out-of-equilibrium processes as soon as only the lightest species remains in the dark sector. In section \ref{sec:dilution} we will discuss in detail the evolution of this stage as well as its consequences for the observable signals of a dark phase transition.

\subsection{Dark scalar effective potential}
In this work we are interested in thermal phase transitions within the dark sector. These transitions can occur when a scalar field has a temperature-dependent vev \cite{Quiros1999,Hindmarsh2020}. The vev of a given field is dictated by the principle of stationary action, where the action is calculated using only the static components of the field. Setting all kinetic terms in the corresponding Lagrangian to zero, the principle of stationary action reduces to the minimization of the field's potential energy density. Since the scalar field of interest is a quantum field in a finite-temperature environment, several corrections have to be added to its tree-level potential. Including finite-temperature effects up to $1$-loop order and daisy diagram contributions to the vacuum energy, the effective potential reads \cite{Delaunay2008}
\begin{align}
V_\text{eff}^\text{1-loop}(\phi,T) = V_\text{tree}(\phi) + V_\text{CW}(\phi)  +V_\text{ct}(\phi) + V_\text{T}(\phi, T) + V_\text{daisy}(\phi, T) \; .
\label{eq:effective_potential}
\end{align}
The first term on the right-hand side is the tree-level potential of the scalar field $\phi$, $V_\text{CW}(\phi) + V_\text{ct}(\phi)$ is the Coleman-Weinberg contribution and the corresponding counterterm, $V_\text{T}$ is the finite-temperature contribution to the vacuum energy density, and $V_\text{daisy}$ encodes the contributions from the resummation of the Matsubara-zero modes of bosonic ring diagrams.

Treating the ultraviolet divergences in $V_\text{CW}$ with dimensional regularization in the $\overline{\text{MS}}$ renormalization scheme and the infrared divergences of the boson modes with vanishing Matsubara frequency in the thermal corrections to $V_\text{tree}$ with the Arnold-Espinoza method~\cite{Arnold1993}, the contributions read
\begin{subequations}
\begin{align}
V_\text{CW}(\phi) &= \sum_x \eta_x \, n_x \,\frac{m_x^4(\phi)}{64\, \pi^2} \bb{\ln \frac{m_x^2(\phi)}{\Lambda^2} - C_x}\; ,\\
V_T(\phi, T) &= \frac{T^4}{2 \,\pi^2} \sum_x \eta_x \,n_x\, J_{\eta_x} \ba{\frac{m_x^2(\phi)}{T^2}}\; ,\\
J_{\eta_x}\ba{z^2} &\equiv \int_{0}^{\infty} \diff y \,  y^2 \ln \bb{1 - \eta_x \, \exp \ba{- \sqrt{y^2 + z^2}}} \; ,\\
V_\text{daisy}(\phi, T) &= - \frac{T}{12 \,\pi} \sum_b n_b^\text{L} \bb{\ba{m^2(\phi) + \Pi(T)}^{3/2}_b - \ba{m^2(\phi)}^{3/2}_b} \; . \label{eq:daisypot}
\end{align}
\end{subequations}
Here, $n_x$ are the degrees of freedom of the fields coupled to $\phi$, $n_b^\text{L}$ are their longitudinal boson components, $\eta_x$ is $+1$ ($-1$) for bosons (fermions), $\Lambda$ is the renormalization scale, which will be set to the tree-level vev $v$ of $\phi$, and $C_x = 3/2$ are the renormalization constants for scalars and fermions, while $C_x = 5/6$ holds for gauge bosons. Goldstone modes have to be counted in addition to the longitudinal gauge boson degrees of freedom (see Ref.~\cite{Delaunay2008}) and the expression $\ba{m^2(\phi) + \Pi(T)}^{3/2}_b$ has to be understood as the $b$-th eigenvalue of the temperature-dependent mass matrix. The functions $\Pi(T)$ denote the hard Debye masses of the longitudinal gauge boson components.

For a quartic tree-level potential $V_\text{tree}(\phi) = - \frac{\mu^2}{2} \phi^2 + \frac{\lambda}{4} \phi^4$, as we will consider in section~\ref{sec:model}, the counterterm potential is given by
\begin{align}
V_\text{ct}(\phi) = - \frac{\delta \mu^2}{2} \phi^2 + \frac{\delta \lambda}{4} \phi^4 \; ,
\end{align}
where the counter-mass $\delta \mu^2$ and the counter-coupling $\delta \lambda$ can be calculated using~\cite{Baker2017}
\begin{subequations}
\begin{align}
\delta \mu^2 &= \left. \bb{\frac{3}{2 \,\phi} \td{V_\text{CW}(\phi)}{\phi} - \frac{1}{2} \tdd{V_\text{CW}(\phi)}{\phi}}\right|_{\phi = \Lambda} \; , \label{eq:counterterms1} \\
\delta \lambda &= \left. \bb{\frac{1}{2 \,\phi^3} \td{V_\text{CW}(\phi)}{\phi} - \frac{1}{2\, \phi^2} \tdd{V_\text{CW}(\phi)}{\phi}}\right|_{\phi = \Lambda} \; . \label{eq:counterterms2}
\end{align}
\end{subequations}

\subsection{First-order phase transitions in the dark sector}
The transition of the real part of the dark Higgs field to different vevs can occur in two different fashions: continuously or discontinuously. In the first case, the global minimum of the effective potential shifts continuously with decreasing temperature, while in the opposite case competing minima in field space occur, to which $\phi$ has to tunnel to minimize its action. As was shown in Ref.~\cite{Linde1980}, the euclidean tunneling action is given by
\begin{align}
S\bb{\phi, T} = \frac{S_3 \bb{\phi, T}}{T} = \frac{1}{T} \int \diff^3 x  \bb{\frac{\ba{\nabla \phi}^2}{2} + V_\text{eff}(\phi, T)} \; , \label{eq:tunnelingaction}
\end{align}
if the field is embedded in a sufficiently hot thermal bath. Imposing stationarity of the action and considering $\text{O}(3)$-symmetric solutions, this yields the so-called bounce equation
\begin{align}
\tdd{\phi}{r} + \frac{2}{r} \td{\phi}{r} = V_\text{eff}^\prime(\phi, T)\;, \label{eq:bounceeq}
\end{align}
with the boundary conditions $\phi(r \rightarrow \infty ) \rightarrow 0$ and $\phi^\prime(r = 0) = 0$. The euclidean distance measure $r = \left|\mathbf{x}\right| = \sqrt{R^2 + c^2 \, t^2}$ can be understood as the radius of a single expanding bubble that reaches luminal bubble wall velocities after having nucleated with an initial radius $R$. 

By solving the bounce equation for a given temperature, one obtains a bubble profile $\phi_T(r)$. Plugging this solution into eq.~\eqref{eq:tunnelingaction}, one obtains the bounce action $S(T) \equiv S[\phi_T, T]$. The bubble nucleation rate per unit volume can now be obtained by computing $\Gamma \simeq T^4 \, e^{-S(T)}$. Comparing this rate with the Hubble rate $H(T)$ at a given temperature yields the nucleation condition $\Gamma(T) \, H^{-4}(T) \overset{!}{=} 1$. Using eqs.~\eqref{eq:Friedmann12} and~\eqref{eq:rhotot} to compute the Hubble parameter, the nucleation criterion thus reads \cite{Breitbach2019}
\begin{align}
\label{eq:nucleation}
	S(T_\text{DS}^\text{n}) \simeq 146 - 2 \ln \ba{\frac{g_{\text{eff},\rho}^\text{tot,n}}{100}} - 4 \ln \ba{\frac{T_\text{DS}^\text{n}}{100 \, \text{GeV}}} \; .
\end{align}
For simplicity, numerical factors of $\mathcal{O}(1)$ for the conversion of the dark sector temperature to a SM temperature are ignored in the second logarithm, here.
This equation can be solved iteratively for the dark sector nucleation temperature $T_\text{DS}^\text{n}$. We emphasize that this procedure is an numerically expensive task, as it requires the computation of the bounce action $S(T_\text{DS})$ at each iteration step. 

For the effective potential introduced above, one can show that the hard thermal loops of the scalar field cancel the thermally induced potential barrier \cite{Breitbach2019}. The same holds for the longitudinally polarized modes of gauge bosons coupled to $\phi$. The transversal modes of coupled gauge bosons however do not obtain thermal masses and can therefore still contribute to the thermally induced barrier. As we are interested in thermally induced first-oder phase transitions, we will in the following consider the case that the scalar is complex and charged under a gauge group. In this way, once the scalar field undergoes spontaneous symmetry breaking, it will give rise to a mass of the gauge boson it is coupled to, with the longitudinal polarization corresponding to the degree of freedom encoded in the angular mode of $\phi$. As one can always project the vev of the complex scalar field to lie on its real axis, the above discussion of the effective potential including only a single scalar component $\phi$ still applies. Since the coupled gauge boson will become massive in the phase transition, we will refer to the complex scalar as a dark Higgs field in the following, in analogy to the SM Higgs field. Accordingly, the massive $\phi$ bosons will be referred to as dark Higgs bosons.

\subsection{Gravitational waves from a dark first-order phase transition}
\label{sec:FOPT}

The emission of GWs in a first-order phase transition is a result of the collision of bubbles in which $\phi$ already obtained its new vev. Additional contributions to the GW spectrum come from the excitation of the primordial plasma during the collision of bubbles in the form of sound waves and magnetohydrodynamic turbulence. The GW spectrum at its emission $\Omega_\text{GW}^\text{em}(f)$ can be calculated as described in eq.~(19) in Ref.~\cite{Breitbach2019}.

The quantity $\alpha$ is a measure of the strength of a first-order phase transition. It is proportional to the amount of energy and pressure liberated in the phase transition, which can be characterized by the difference $\Delta \theta$ of the trace of the energy momentum tensor between the broken and unbroken phase \cite{Caprini2019}, that is 
\begin{align}
\Delta \theta = \left. \left(- \Delta V_\text{eff}(T_\text{DS}) + \frac{1}{4} T_\text{DS} \pd{\Delta V_\text{eff}(T_\text{DS})}{T_\text{DS}}\right)\right|_{T_\text{DS} = T_\text{DS}^\text{n}} > 0 \; . \label{eq:epsilon}
\end{align}
Here, $\Delta V_\text{eff}(T_\text{DS}) < 0$ denotes the difference in potential energy density of the two minima of $V_\text{eff}$, between which $\phi$ tunnels in the phase transition. To receive the dimensionless parameter that quantifies the strength of the phase transition, $\Delta \theta$ is normalized to the total energy density of the surrounding plasma of relativistic species $\rho_\text{tot}^\text{n} = \frac{\pi^2}{30} \, g_{\text{eff},\rho}^\text{tot,n} \, \ba{T_\text{SM}^\text{n}}^4$, such that
\begin{align}
\alpha \equiv \frac{\Delta \theta}{\rho_\text{tot}^\text{n}} \; . \label{eq:alpha}
\end{align}
As $\rho_\text{tot}^\text{n}$ scales approximately with $\xi_\text{n}^{-4}$ for a fixed dark sector nucleation temperature $T_\text{DS}^\text{n}$ and assuming $g_{\text{eff},\rho}^\text{SM,n} \gg g_{\text{eff},\rho}^\text{DS,n}$, the transition strength scales as $\alpha \propto \xi_\text{n}^4$ as was first shown in Ref.~\cite{Breitbach2019}. This is the reason why first-order phase transitions from dark sectors already slightly hotter than the SM bath can potentially emit strong GW signals.

Another important quantity in our analysis arises when one instead normalizes $\Delta \theta$ to the energy density of only the dark sector species, that is
\begin{align}
	\alpha_\text{DS} \equiv \frac{\Delta \theta}{\rho_\text{DS}^\text{n}} \; .
\end{align}
This strength parameter will be used to calculate the energy budget of the contributions from the individual GW sources, which should only depend on the hydrodynamics of the dark sector after the phase transition but not on the decoupled SM bath. The parameter $\alpha_\text{DS}$ is therefore independent of the temperature ratio $\xi_\text{n}$ for a fixed dark sector nucleation temperature $T_\text{DS}^\text{n}$ as opposed to the transition strength $\alpha$.

The inverse time-scale $\beta/H$ of the phase transition can be computed using the derivative of the bounce action $S(T_\text{DS})$ at the nucleation:
\begin{align}
\frac{\beta}{H} \equiv T_\text{DS}^\text{n} \left.\td{S(T_\text{DS})}{T_\text{DS}} \right|_{T_\text{DS}=T_\text{DS}^\text{n}}. \label{eq:betaH}
\end{align}
A fast transition happens on a short time scale and thus leads to a large $\beta/H$, which damps the resulting spectrum. This damping is due to the almost simultaneous production of many bubbles in a fast transition, which will collide while still being relatively small. In the opposite case of a slow transition, the bubble nucleation rate is low, leading to the eventual collision of larger, more energetic bubbles. Since the small bubbles after a fast transition collide more frequently than in the opposite case, the corresponding spectrum will have its peak at a higher frequency. The ratio $\beta/H$ is almost independent of $\xi_\text{n}$, as was shown in Ref.~\cite{Breitbach2019}.

The bubble wall velocity $v_\text{w}$ is the most intricate parameter, since its calculation requires knowledge of the diverse, highly model-dependent particle processes that can happen at the accelerating bubble wall, see e.\,g.\ Refs.~\cite{Baker2019,Chway2019,Hoche:2020ysm,Azatov2020,Azatov2021}. In general, the collision of particles in the plasma with an expanding bubble exerts a non-negligible pressure on its moving wall. Additionally, next to the mere change of momentum of particles being reflected, there is also an additional friction term due to transition radiation by gauge bosons, which is likely to dominate over the friction from particles colliding with the wall \cite{Bodeker2017}. A detailed analysis of the processes happening at the bubble wall requires the solution of Boltzmann-like equations and is a subject of current research. However, for sufficiently strong first-order phase transitions, the bubble walls will quickly reach luminal velocities. Since strong phase transitions are favorable for detectable stochastic GW backgrounds, we will focus on very strong transitions for which $v_\text{w} \sim 1$, and neglect the details of the bubble wall dynamics.

If $\alpha_\text{DS}$ exceeds a certain threshold strength $\alpha_\infty$, the bubble walls can accelerate continuously (the ``runaway regime''), while in the opposite case there will be a terminal velocity (the ``non-runaway regime''). One can show that the friction exerted by particles getting (more) massive in a first-order phase transition is approximately given by~\cite{Espinosa:2010hh}
\begin{align}
P_\text{fric} \approx \Delta V_\text{T} \approx \ba{T_\text{DS}^\text{n}}^2 \bb{\sum_b \frac{n_b}{24} \, \Delta m_b^2(\phi) + \sum_f \frac{n_f}{48} \, \Delta m_f^2(\phi)} \; .
\label{eq:Pfric}
\end{align}
The condition $\epsilon > P_\text{fric}$ for runaway bubbles is thus equivalent to $\alpha_\text{DS} > \alpha_\infty \equiv P_\text{fric} / \rho_\text{DS}(T_\text{DS}^\text{n})$, which will be used as a definition for the threshold transition strength. Note that the expression in eq.~\eqref{eq:Pfric} is meant to sum over all particles gaining masses in the phase transition and thus excludes Goldstone bosons~\cite{Caprini2015}.

The efficiency factors $\kappa$ of the three contributions entering $\Omega_\text{GW}^\text{em}(f)$ depend on both $\alpha_\text{DS}$ as well as the coupling between the plasma and the bubble wall. If the bubble walls have a terminal velocity, the latent heat of the transition is rather converted into kinetic energy of the plasma than used to accelerate the bubble walls. Consequently, the contribution from bubble collisions to the GW signal are negligible in the non-runaway bubble scenario, that is $\kappa_\phi = 0$. The efficiency factor $\kappa_\text{sw}$ of sound wave contributions then follows a function $\kappa(\alpha_\text{DS})$ that can be approximated for luminal bubble wall velocities $v_\text{w} \sim 1$ as \cite{Caprini2015} 
\begin{align}
\kappa(\alpha_\text{DS}) \approx \frac{\alpha_\text{DS}}{0.73 + 0.083 \, \sqrt{\alpha_\text{DS}} + \alpha_\text{DS}}\;.
\end{align}
Conversely, when the bubble walls can accelerate continuously, bubble collisions are non-negligible sources of GWs and contribute with $\kappa_\phi = 1 - \alpha_\infty/ \alpha_\text{DS}$. The efficiency of sound wave contributions to the GW signal in the runaway regime then reads $\kappa_\text{sw} = \kappa(\alpha_\infty) \,\alpha_\infty / \alpha_\text{DS}$. In either case, a fraction $\epsilon_\text{turb}$ of the bulk motion energy is converted into turbulence, such that $\kappa_\text{turb} = \epsilon_\text{turb} \, \kappa_\text{sw}$. Following Ref.~\cite{Breitbach2019} we employ the optimistic estimate $\epsilon_\text{turb} \simeq 10\,\%$.

An accurate prediction of the spectrum of GWs generated in a first-order phase transition in the early universe for a given model can be very challenging, requiring advanced methods for the calculation of the effective potential~\cite{Croon:2020cgk,Gould:2021oba}, for the onset and duration of the phase transition, as well as for a description of bubble walls~\cite{Guo:2021qcq}.
The focus of the present work is however not on the detailed dynamics of the phase transition itself, but on the cosmological evolution of the dark sector that gives rise to such a phase transition subsequent to the generation of a GW signal. We will therefore limit ourselves to the simplified calculation of the stochastic GW background shown above with the understanding that all the effects discussed in the remainder of this work would equally apply to a more refined approach. We note however that the most important uncertainties of our calculations are either negligible or lead to a conservative estimate of the resulting signals of the analyzed first-order phase transitions as we explain in the following.

As was described in Ref.~\cite{Caprini2019}, it is not clear whether the equations used to compute the contributions to $\Omega^\text{em}_\text{GW}(f)$ from sound waves and magnetohydrodynamic turbulence in the primordial plasma can be applied to the case of two decoupled sectors. In our analysis, we will focus on transitions deep in the runaway bubble regime for which $\alpha_\text{DS} \gg \alpha_\infty$, such that $\kappa_\phi \gg \kappa_\text{sw}$ and $\kappa_\phi \gg \kappa_\text{turb}$. Thus, virtually no latent heat gets transferred to bulk plasma motion such that bubble collisions provide the dominant source of the emitted stochastic GW background. Therefore, the uncertainties connected with the sound wave and turbulence production of a stochastic GW background are negligible for the presented analysis.

The description of contributions from bubble collisions to $\Omega_\text{GW}^\text{em}(f)$ however relies heavily on previous semi-analytical work that utilized the envelope approximation. In this approach, it is assumed that the scalar field's stress-energy is located in an infinitesimally thin shell around the bubble wall, which vanishes when two bubbles collide. Lately, this approximation was shown to only yield slightly larger signal strengths than predicted by fully numerical simulations of bubble collisions for vacuum transitions in the runaway bubble regime and to outperform the alternative bulk flow model \cite{Cutting2020}. However, Refs.~\cite{Ellis:2019oqb,Lewicki:2020jiv,Lewicki:2020azd} find substantial deviations from the envelope approximation for strongly supercooled phase transitions. Nevertheless, these uncertainties do not affect the main findings of our work and are therefore neglected in the following.

A general review of more refined computational approaches than the one used in this work can be found in Ref.~\cite{Guo:2021qcq}. Most importantly, we used the nucleation temperature as a reference scale to define thermodynamic quantities. New findings suggest to rather employ the ``percolation temperature'' instead, corresponding to the point in time when a significant fraction of the Universe is already filled with bubbles of the broken phase. This especially makes a difference for very strong transitions with $\alpha \gtrsim 1$, as in these cases the percolation temperature is significantly lower than the nucleation temperature \cite{Wang2020}. Then, also the energy density $\rho_\text{tot}$ that is used as a normalization in the definition of $\alpha$ in eq.~\eqref{eq:alpha} is lower, such that $\alpha$ increases effectively. Our approach hence yields a good approximation for transition strengths up to about $1$ and underestimates the expected signal strength for even stronger transitions. 

\paragraph{Redshift of the GW background after its emission.}
After its generation, the stochastic GW background propagates freely and undisturbed until today, effectively being a form of dark radiation.\footnote{We note that there are a number of exceptions to this statement. For example, there is an enhancement of modes that enter the horizon during matter domination~\cite{Barenboim:2016mjm,Ellis:2020nnr}. For a discussion of deviations from our approximation we refer to Ref.~\cite{Hook2020}.} The expansion of the Universe, however, redshifts both its amplitude and its frequency such that today's power spectrum, $\Omega_\text{GW}(f)$, can be expressed as
\begin{align}
\Omega_\text{GW}(f) = \mathcal{R}\, \Omega_\text{GW}^\text{em}\ba{\frac{a_0}{a_\text{n}} f} \; , \label{eq:Omegaredshift}
\end{align}
where $f$ denotes the spectrum's frequency in today's units, which is shifted from its value at nucleation by multiplication with the scale factor ratio $a_0 / a_\text{n}$, where $a_\text{n}$ ($a_0$) is the scale factor at nucleation (today) \cite{Breitbach2019}. The amplitude of the spectrum redshifts like $a^{-4} \,H^{-2}$, since the energy density of radiation scales with $a^{-4}$, while the critical energy density $\rho_\text{c} = 3 \, m_\text{Pl}^2 \, H^2$ with which $\rho_\text{GW}$ has been normalized to obtain $\Omega_\text{GW}$ scales with $H^{2}$. The prefactor $\mathcal{R}$ is thus defined as
\begin{align}
\mathcal{R} \equiv \ba{\frac{a_\text{n}}{a_0}}^4 \ba{\frac{H_\text{n}}{H_0}}^2. \label{eq:fancyR}
\end{align}		
In the following chapter \ref{sec:dilution}, we will see how the dilution effect by the out-of-equilibrium decay of a dark sector into SM particles can contribute to the ratio $a_0 / a_\text{n}$ and the quantity $\mathcal{R}$.

\section{Evolution and decay of a hot dark sector}\label{sec:dilution}

In this section we consider the evolution of the dark sector \emph{after} the phase transition has ended and a stochastic GW background has been produced. Since we are interested in temperature ratios $\xi > 1$, large amounts of energy are stored in the dark sector and need to be transferred to the visible sector in order to satisfy observational constraints. This transfer of energy implies that the entropy in each sector is no longer conserved, i.\,e.\ it is no longer possible to directly calculate the scale factor $a_\mathrm{n}$ of bubble nucleation in terms of the SM temperature $T^\mathrm{n}_\text{SM}$. As a result, the present-day GW signal depends not only on the details of the phase transition itself, but also on the subsequent energy transfer from the dark to the visible sector.

In principle, there are a number of different ways for depleting the energy density of the dark sector. Here, we will focus on the case that the lightest particle in the dark sector (henceforth referred to as the mediator) is unstable and eventually decays into SM particles. If the mediator has a very long lifetime, it will come to completely dominate the energy density of the universe leading to an early period of matter domination (or, if number-changing processes are efficient, of cannibal domination). When these particles eventually decay,  the resulting entropy injection into the SM sector may then lead to a significant dilution of any previously produced GW signal. We define the dilution factor as the ratio of the entropy in the SM sector before and after the mediator decays:
\begin{equation}
	D_\text{SM} \equiv \frac{S_\text{SM}^\text{after}}{S_\text{SM}^\text{before}} \; . \label{eq:dilutionfactorD}
\end{equation}

Most of the discussion in this section is very general in the sense that the dilution factors that we calculate apply to a wide range of dark sectors and any type of GW signal (or, in fact, any form of fully decoupled matter or radiation). Nevertheless, we will assume for concreteness that the mediator is a scalar boson, for example a dark Higgs boson originating from spontaneous symmetry breaking in the dark sector. In this case, $3 \to 2$ processes play an important role in the evolution of the dark sector energy density once the temperature of the dark sector drops below the mass of the lightest state.

For simplicity, we assume that decays of the mediator happen after this particle has become non-relativistic. This assumption implies in particular that all heavier dark sector particles have annihilated away and transferred their energy to the lightest states. We denote the temperature at which the heavier states decouple by $T_\text{DS}^\text{cd}$. Moreover, if the mediator particles are non-relativistic when they decay, we can neglect effects related to inverse decays from SM states. On the other hand, the requirement that BBN proceeds as in standard cosmology places an upper bound on the lifetime of the lightest dark sector state, which is approximately given by $\tau < 1 \, \mathrm{s}$~\cite{Hufnagel2020}.

We will begin our discussion by considering the evolution of a dark sector away from thermal equilibrium in section \ref{subsec:mediatorevo}. The equations describing the mediator decays will be derived and solved numerically in section \ref{subsec:entropyinjection}. The overall effect of these decays on GW signals will be investigated in section \ref{subsec:dilution}.

\subsection{Evolution of the mediator energy density} \label{subsec:mediatorevo}

Once the mediator is the only particle species remaining in the dark sector, the only mechanisms that can change its energy density are the expansion of the universe, possible number-changing processes and the eventual decays of the mediator into SM particles. We will first consider the case that number-changing processes are negligible and then extend our discussion to include mediator cannibalism.

In principle, the Boltzmann equation for the phase space density of a decaying mediator with no other interactions can be directly integrated given the time dependence of the scale factor $a(t)$. The energy density $\rho_\text{med}(t)$ is then obtained by integrating the distribution function over momentum space~\cite{Hufnagel2020}. Since this procedure is numerically rather expensive, we present below a simple approximation that can be used to describe the thermal history of the mediator species from its chemical decoupling until its decay. 

Following the approach presented in Ref.~\cite{Cirelli2018} and extending it to allow for relativistic mediators, we can approximately write
\begin{align}
\td{\rho_\text{med}}{t} = - 3 \, \zeta(t) \, H(t) \, \rho_\text{med}(t) - \frac{\rho_\text{med}(t)}{\tau} \; , \label{eq:Boltzmannapprox}
\end{align}
where $\tau$ is the lifetime of the mediator species and $\zeta(t)$ is an appropriate function (to be discussed in more detail below) such that $\zeta(t) = 4/3$ for relativistic mediators and $\zeta(t) = 1$ for non-relativistic mediators. We point out that eq.~\eqref{eq:Boltzmannapprox} implicitly assumes that the decays of the mediator only become relevant after the mediator has become non-relativistic, because we have not included a temperature-dependent Lorentz factor to account for time dilation in the final term~\cite{Hufnagel2020}. 

As initial condition we assume that $\rho_\text{med}$ is given by an equilibrium distribution with vanishing chemical potential at sufficiently early times. In practice, we start our calculation at the time when all heavier states in the dark sector have decoupled, which we denote by $t_\mathrm{cd}$ (for chemical decoupling), assuming $t_\mathrm{cd} \ll \tau$.
To simplify notation, we introduce the dimensionless quantities 
\begin{align}
\theta & \equiv t/\tau \, , \\
\bar{a} & \equiv a / a_\text{cd}\;,
\end{align}
and denote derivatives with respect to $\theta$ by a prime. Eq.~(\ref{eq:Boltzmannapprox}) then becomes

\begin{align}
\rho_\text{med}^\prime(\theta) = - 3 \, \zeta(\theta) \, \frac{\bar{a}^\prime}{\bar{a}} \, \rho_\text{med}(\theta) - \rho_\text{med}(\theta) \; .
\label{eq:non-reldecay}
\end{align}

\begin{figure}[t]
\centering
\includegraphics[width=0.65\linewidth]{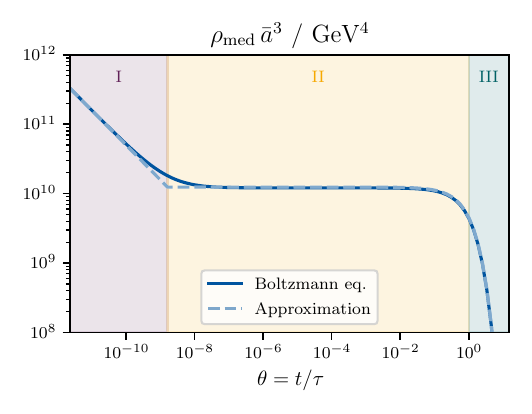}
\caption{Time evolution of the comoving energy density $\rho_\text{med}\, \bar{a}^3$ of a mediator decaying to SM particles for the case that number-changing processes are negligible. The solid line indicates the evolution obtained by integrating the full Boltzmann equation, whereas the dashed line shows the result from integrating the approximation in eq.~\eqref{eq:non-reldecay}.  For concreteness, we have taken the mediator mass to be $m_\text{med} = 100 \, \mathrm{GeV}$ and assumed that the decoupling of other dark sector species from the mediator occurs at $T_\text{DS}^\text{cd} = 1 \, \text{TeV}$. Setting $\xi_\text{cd} = 1$ and $\tau = 0.1 \, \text{s}$ results in the dimensionless time parameters $\theta_\text{cd} = 2.3 \cdot 10^{-12}$ for the chemical decoupling and $\theta_\text{nr} = 1.7 \cdot 10^{-9}$ for the mediator becoming non-relativistic.}
\label{fig:rhomed}
\end{figure}

In figure~\ref{fig:rhomed}, we compare the solution of this equation to the result obtained by integrating the full Boltzmann equation, assuming radiation domination for simplicity. We find that a good overall agreement between the two curves is achieved when performing the transition from $\zeta = 4/3$ to $\zeta = 1$ when $T_\text{DS} \approx 0.38 \, m_\text{med}$, which can be written as\footnote{
	We point out that $\theta_\text{nr}$ does not depend on $T_\text{DS}^\text{cd}$. This can be seen explicitly in the cancellation of $\ba{T_\text{DS}^\text{cd}}^2$ in $\theta_\text{nr} \propto \theta_\text{cd} \ba{T_\text{DS}^\text{cd}}^2 \propto H_\text{cd}^{-1} \, \ba{T_\text{DS}^\text{cd}}^2$, where $H_\text{cd} \propto \ba{T_\text{DS}^\text{cd}}^2$ is the Hubble parameter at the chemical decoupling.
}

$\theta \approx \theta_\mathrm{nr} \equiv 7.0 \, \theta_\text{cd} \ba{T_\text{DS}^\text{cd} / m_\text{med}}^2$. In this approximation, the mediator behaves as a relativistic particle species with $\rho_\text{med} \propto a^{-4}$ for $\theta < \theta_\mathrm{nr}$, while for $\theta > \theta_\text{nr}$ it behaves as non-relativistic matter and its energy density scales as $\rho_\text{med} \propto a^{-3}$. For $\theta \gtrsim 1$, mediator decays become relevant such that the energy density decreases exponentially as $\rho_\text{med} \, a^3 \propto \exp (- \theta)$. The assumption that the mediator is non-relativistic when it decays therefore translates to the requirement $\theta_\mathrm{nr} \ll 1$. In this case, a precise description of the mediator energy density around $\theta_\text{nr}$ is irrelevant for its subsequent evolution and hence the approximation introduced above is sufficient for our purposes.

So far, we have assumed that the only processes changing the comoving number density are decays of the mediator into SM particles. As was first argued in Ref.~\cite{Carlson:1992fn}, however, this description is incomplete, because a secluded particle species can perform number-changing processes like $3 \rightarrow2$ or $4 \rightarrow2$, thereby reducing its comoving number density while conserving its entropy. This leads to an unusual relationship between the energy density and the scale factor until the number-changing processes become inefficient. Since in this process the species consumes itself to keep warm (i.\,e.\ to prevent becoming non-relativistic), it is casually referred to as ``cannibalism'' \cite{Farina:2016llk}.\footnote{We point out that, although the temperature decreases more slowly in a cannibalistic dark sector with vanishing chemical potential than in a non-interacting dark sector with non-zero chemical potential, the energy density decreases more rapidly. As we will see in section~\ref{subsubsec:derivingODEs}, cannibalism therefore reduces the dilution factor due to entropy injection.}

If number-changing processes are efficient, i.\,e.\ their rate exceeds the Hubble rate, the chemical potential of the particle species vanishes. This is typically the case at sufficiently early times and high temperatures when the number densities are large. In this case the equilibrium energy density $\rho_\text{med}$ and the equilibrium entropy density $s_\text{med}$ only depend on $z_\text{med} \equiv m_\text{med} / T_\text{DS}$. By eliminating $z_\text{med}$ we can therefore obtain a function $\bar{s}_\text{med}(\bar{\rho}_\text{med})$, where $\bar{s}_\text{med} \equiv 2 \, \pi^2 \, s_\text{med}  / (g_\text{med} \, T_\text{DS}^3)$ and $\bar{\rho}_\text{med} \equiv 2 \, \pi^2 \, \rho_\text{med} / (g_\text{med} \, T_\text{DS}^4)$. 

Since we know that, as long as the decay rate of the mediator is negligible, the dark sector entropy is conserved ($s_\text{med} \, a^3 = \text{const}$), we can use this function to calculate the evolution of the energy density of the mediator species:
			\begin{align}
				\dot{\rho}_\text{med} &= - 3 \, \td{\ln \rho_\text{med}}{\ln s_\text{med}} \, H(t) \, \rho_\text{med}(t) = - 3 \,  \td{\ln \bar{\rho}}{\ln \bar{s}} \, H(t) \, \rho_\text{med}(t) \; .\label{eq:rhomedcannibalism}
			\end{align}
We note that the function $\td{\ln \bar{\rho}}{\ln \bar{s}}(\rho_\text{med})$ is close to $4/3$ for large energy densities, corresponding to high temperatures and relativistic species, and approaches $1$ for low energy densities $\rho_\text{med}$, corresponding to non-relativistic species. Hence, for the case that number-changing processes are efficient, $\td{\ln \bar{\rho}}{\ln \bar{s}}$ replaces the function $\zeta(\theta)$ introduced above to describe the transition from the relativistic to the non-relativistic scaling.

As the number density of the mediator species decreases, number-changing processes eventually become inefficient and the chemical potential $\mu_\text{med}$ can no longer be neglected. This transition is typically quite sudden, meaning that number-changing processes are either sufficient to keep the mediators in chemical equilibrium, or completely negligible~\cite{Heimersheim:2020aoc}. Hence, as soon as the rate of number-changing processes $\Gamma_\text{nc}$ drops below the Hubble rate, we can revert to the description for non-interacting mediators from above.
To combine both of these phases, we can therefore define
\begin{align}
\zeta(\theta) = \begin{cases}
\td{\ln \bar{\rho}}{\ln \bar{s}} \ba{\rho_\text{med}} & \quad \text{for} \quad \Gamma_\text{nc}(\theta) \ge H(\theta) \\
\frac{4}{3} &  \quad \text{for} \quad \Gamma_\text{nc}(\theta) < H(\theta), \quad \theta < \theta_\text{nr}\\
1&  \quad \text{for} \quad \Gamma_\text{nc}(\theta) < H(\theta), \quad   \theta \ge \theta_\text{nr}
\end{cases} \; , \label{eq:zeta}
\end{align}
which can be used in eq.~\eqref{eq:non-reldecay} to include the effect of cannibalism.

The only remaining task is then to calculate $\Gamma_\text{nc}$ for a given particle physics model. In the absence of a stabilising symmetry, the dominant contribution typically arises from the $3 \rightarrow 2$ rate $\Gamma_{32} \approx \langle \sigma_{32} \, v^2 \rangle \,  n_\text{med}^2 \approx \langle \sigma_{32} \, v^2 \rangle \,  \rho_\text{med}^2 / m_\text{med}^2$ \cite{Farina:2016llk}, where the thermally averaged cross section of the $3 \rightarrow2$ process can be written as
\begin{align}
\langle \sigma_{32}\,  v^2 \rangle = \frac{25 \sqrt{5} \, \pi^2}{5184} \; \frac{\alpha_{32}^3}{m_\text{med}^5}+ \mathcal{O}\ba{\frac{T_\text{DS}}{m_\text{med}}}\;,
\end{align}
for a scalar mediator. If the scalar potential is written as $V(\phi) = \frac{m_\mathrm{med}^2}{2} \, \phi^2 + \frac{\kappa_3}{3!} \,\phi^3 + \frac{\kappa_4}{4!} \,\phi^4$, the effective $3\rightarrow2$ coupling is given by \cite{Erickcek:2020wzd,Farina:2016llk}
\begin{align}
	\ba{4 \, \pi \, \alpha_{32}}^3 \equiv \ba{\frac{\kappa_3}{m_\mathrm{med}}}^2 \bb{\ba{\frac{\kappa_3}{m_\mathrm{med}}}^2 + 3 \, \kappa_4}^2.
\end{align}
If the mediator acquires its mass through the spontaneous breaking of a symmetric potential with quartic interactions,  $\kappa_3 = \sqrt{3\, \kappa_4} \, m_\mathrm{med}$ holds after the phase transition.

\begin{figure}[t]
\centering
\includegraphics[width=.85\linewidth]{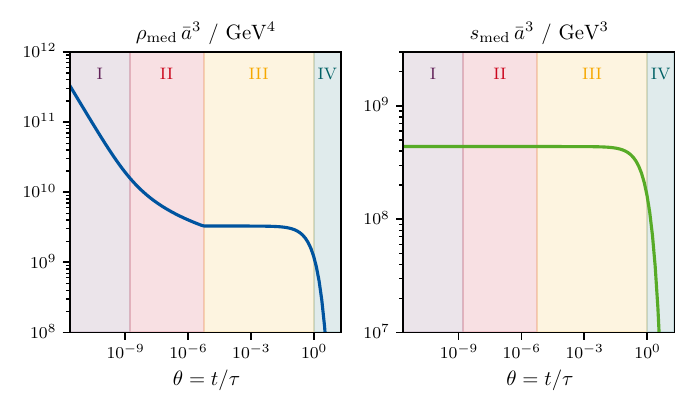}
\caption{\textit{Left}: Plot of the time evolution of a dark Higgs mediator species with the same specifications as in figure~\ref{fig:rhomed}, but with an intermediate phase of cannibalism, characterized by an effective $3 \to 2$ coupling $\alpha_{32} = 0.1$. In the area shaded in violet (I), the mediator species is still relativistic, such that $\rho_\text{med} \propto \bar{a}^{-4}$, while in the red area (II), number-changing processes lead to a decrease of $\rho_\text{med}  \, \bar{a}^3 \propto 1 / \ln \bar{a}$ with the normalized scale factor $\bar{a}$. In the area shaded yellow (III), the mediator starts to decay non-relativistically, i.\,e.\ $\rho_\text{med} \,  \bar{a}^3 \propto e^{- \theta}$, and in the blue shaded area (IV), we have $\theta > 1$, indicating the decay of the mediator becoming the dominant effect. \textit{Right}: The entropy $s_\text{med} \, \bar{a}^3$ is conserved until the mediator species decays.}
\label{fig:cannibalism}
\end{figure}

The resulting time evolution of $\rho_\text{med} \, \bar{a}^3$ and $s_\text{med} \, \bar{a}^3$ is shown in figure~\ref{fig:cannibalism}. In the violet shaded area (phase I) and the red shaded area (phase II), $\Gamma_\text{nc} \ge H$ holds, such that number-changing processes are efficient, the chemical potential is negligible and $\zeta(\theta)$ follows the gradual decrease from $4/3$ to $1$ as described above. The transition between phase I and II corresponds to $\theta = \theta_\mathrm{nr}$. For larger $\theta$, the mediator energy density scales as $\rho_\text{med} \, \bar{a}^3 \propto 1 / \ln \bar{a}$\ \cite{Carlson:1992fn}. The end of the cannibalism period, i.\,e.\ the transition from phase II to III is given by the condition $\Gamma_\text{nc} = H$. From this point onward, the expected behavior for a non-relativistically decaying species, $\rho_\text{med} \, \propto a^{-3} \, \exp(- \theta)$, is recovered. As one can see on the right-hand side of figure~\ref{fig:cannibalism}, entropy is  conserved throughout the first two phases and only decreases when the mediators start decaying. 

Comparing the left panel of figure~\ref{fig:cannibalism} to figure~\ref{fig:rhomed}, for which the same benchmark point (but without including number-changing processes) was considered, shows that a cannibalism phase can substantially reduce the energy density stored in the dark sector. Eventually, this reduces the amount of energy injected into the SM bath and hence results in less reheating of the SM bath, as will be shown in the next section.

\subsection{Entropy injection into the Standard Model} \label{subsec:entropyinjection}

For the discussion above we have assumed that the scale factor is proportional to $\sqrt{t}$, corresponding to a radiation dominated era. However, this is not necessarily the case in our set-up, because the non-relativistic mediators can come to dominate the energy density of the universe, leading to an early era of matter domination. The period only ends when the mediators decay, leading to the injection of a considerable amount of entropy into the SM bath~\cite{Scherrer1985,Kolb1990}. It is well known that such an entropy injection can have a profound impact on the abundance of a frozen-out dark matter component~\cite{Cirelli2018}. As we will see below, the same is true for a stochastic GW background produced before the era of matter domination. Hence, it is essential for our purposes to obtain an accurate description of the relevant effects.

\subsubsection{Differential equations governing entropy injection} \label{subsubsec:derivingODEs}

As we are interested in the process of entropy transfer from one sector to another through an exchange of energy, we need to simultaneously consider the evolution of the SM and dark sector energy densities, the SM degrees of freedom and the scale factor. If the energy density of the mediators $\rho_\text{med}(\theta)$ is non-negligible compared to the energy density of SM radiation $\rho_\text{rad}$, the first Friedmann equation becomes
\begin{equation}
\bar{a}^\prime = \frac{\bar{a}}{\theta_\text{H}} \sqrt{r + \frac{r_\text{rad}^\text{cd}}{\bar{a}^4} \,  \frac{\mathcal{S}}{\mathcal{G}^{1/3}}} \; ,
\label{eq:ODE_1}
\end{equation}
where we have introduced the dimensionless variable $r(\theta) \equiv \rho_\text{med}(\theta) / \rho_\text{med}^\text{cd}$ and the constants $\theta_\text{H} \equiv \sqrt{3 \, \Mp^2 / (\tau^2 \, \rho_\text{med}^\text{cd})}$ and $r_\text{rad}^\text{cd} \equiv \rho_\text{rad}^\text{cd} / \rho_\text{med}^\text{cd}$. Furthermore, we have introduced the functions
\begin{align*}
\mathcal{G}(\theta) \equiv \frac{\gsSM(\theta)}{\gsSMcd}\; ,&&
\mathcal{S}(\theta) \equiv \ba{\frac{S_\text{SM}(\theta)}{S_\text{SM}^\text{cd}}}^{4/3}\;,
\end{align*}
which encode the change in the SM entropy degrees of freedom and the total entropy of the SM sector, respectively. Note that we have implicitly assumed that any other form of non-relativistic matter gives a negligible contribution to the energy density and that $\grhoSM (\theta) \approx \gsSM (\theta)$, which are both excellent approximation for the temperature range that we will be interested in.

As discussed above, the evolution of the mediator energy density is given by eq.~\eqref{eq:non-reldecay}, which can be written as
\begin{equation}
r^\prime = - r - 3 \, \frac{\bar{a}^\prime}{\bar{a}} \, \zeta \, r \label{eq:ODE_2} \; .
\end{equation}
The change of SM entropy due to mediator decays is directly proportional to the mediator energy density and is given by
\begin{equation}
\mathcal{S}^\prime = \frac{r \, \bar{a}^4}{r_\text{rad}^\text{cd}} \, \mathcal{G}^{1/3} \label{eq:ODE_3} \; . 
\end{equation}
Using the relation
\begin{equation}
T_\text{SM}(\theta) = \frac{a}{a_\text{cd}} \ba{\frac{S_\text{SM}}{S_\text{SM}^\text{cd}}}^{1/3} \ba{\frac{\gsSMcd}{\gsSM}}^{1/3} T_\text{SM}^\text{cd} = \frac{\bar{a} \, \mathcal{S}^{1/4}}{\mathcal{G}^{1/3}} T_\text{SM}^\text{cd} \;, \label{eq:TSM}
\end{equation}
we can write the change of the SM degrees of freedom as
\begin{equation}
\mathcal{G}^\prime = - \frac{3}{4} \, \frac{T_\text{SM}^\text{cd} \, \mathcal{G} \, \hat{\mathcal{G}}}{\mathcal{S}^{3/4} \, \bar{a}} \, \frac{4 \, \mathcal{S} \, \bar{a}^\prime - \mathcal{S}^\prime  \, \bar{a}}{T_\text{SM}^\text{cd} \, \hat{\mathcal{G}} \,  \mathcal{S}^{1/4} + 3 \, \mathcal{G}^{4/3} \bar{a}} \; , \label{eq:ODE_4}
\end{equation}
where we have introduced
\begin{align}
\hat{\mathcal{G}}(\theta) = \td{}{T_\text{SM}} \left.\bb{\frac{\gsSM(T_\text{SM})}{\gsSMcd}}\right|_{T_\text{SM}(\theta)} \; ,	\label{eq:Ggammaprime}
\end{align}
see appendix~\ref{app:derivingODEs} for details. Note that, equivalently, the latter equation could be replaced by a differential equation for ${T_\text{SM}}^\prime$.

The differential eqs.~\eqref{eq:ODE_1}--\eqref{eq:ODE_4} can be easily solved with initial conditions given by $\bar{a}_\text{cd} = \mathcal{S}_\text{cd} = r_\text{cd} = \mathcal{G}_\text{cd} = 1$. 
The actual particle physics properties of the dark sector are hidden in the various quantities defined above, specifically $r_\text{rad}^\text{cd}$, $T_\text{SM}^\text{cd}$ and $\theta_\text{H}$, as well as $\theta_\text{nr}$ and $\alpha_{32}$, which enter in the definition of $\zeta(\theta)$ in eq.~\eqref{eq:zeta}. The evolution of the mediator species as well as the scale factor and the SM entropy ratio generated during the decay are therefore fully described by these five parameters.

For a more intuitive interpretation of our results, we would like to express the parameters $r_\text{rad}^\text{cd}$, $\theta_\text{H}$ and $\theta_\text{nr}$ in terms of the mediator lifetime $\tau$, the mediator mass $m_\text{med}$, the temperature ratio $\xi_\text{cd}$ as well as the temperature $T_\text{SM}^\text{cd}$. To do so, we calculate the initial mediator energy density $\rho_\text{med}^\text{cd}$ by setting $T_\text{DS}^\text{cd} = \xi_\text{cd} \, T_\text{SM}^\text{cd}$  and assuming that the mediator has an equilibrium distribution with vanishing chemical potential at chemical decoupling. As mentioned above, we limit ourselves to the case that the degrees of freedom of the mediator are given by $g_\text{med} = 1$. The initial ratio of energy densities $r_\text{rad}^\text{cd}$ can then be calculated directly from $T_\text{SM}^\text{cd}$, using the appropriate number of relativistic degrees of freedom. 
The age of the Universe at chemical decoupling $t_\text{cd}$ in units of $\tau$ is given by $\theta_\text{cd} = \ba{2 \, H_\text{cd} \, \tau}^{-1}$, where $H_\text{cd}^2 = \frac{\pi^2}{90} \, g_{\text{eff},\rho}^\text{tot,cd}  \, \ba{T_\text{SM}^\text{cd}}^4 / m_\text{Pl}^2$ denotes the Hubble parameter at chemical decoupling. Knowing $\theta_\text{cd}$, we are then able to determine both time parameters $\theta_\text{nr} \equiv 7.0 \, \theta_\text{cd} \ba{T_\text{DS}^\text{cd} / m_\text{med}}^2$ and $\theta_\text{H} \equiv \sqrt{3 \,\Mp^2 \,/ (\tau^2 \,\rho_\text{med}^\text{cd})}$.

\subsubsection{Numerical solution}

\begin{figure}
\centering
\includegraphics[width=.9\linewidth]{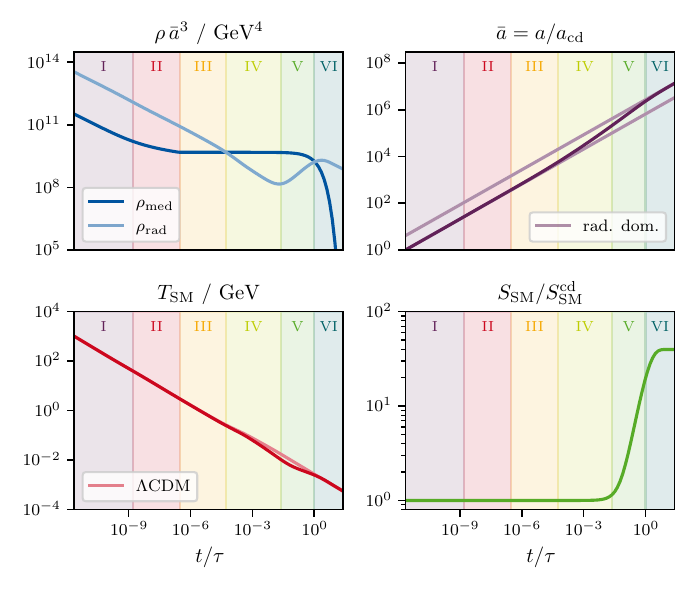}
\caption{
Time evolution of the comoving energy densities $\rho \, \bar{a}^3$ of the mediator species and the SM radiation (top-left), the normalized scale factor $\bar{a}$ (top-right), the temperature $T_\text{SM}$ of the SM bath (bottom-left), as well as its entropy $S_\text{SM}/ S_\text{SM}^\text{cd}$ (bottom-right). The evolution can be divided into the following phases: Relativistic mediator (I), cannibalism (II), non-relativistic mediator (III), early matter domination (IV), entropy injection (V), and decay (VI). See text for details.}
\label{fig:dilution}
\end{figure}

Let us consider an example to highlight the various evolutionary stages encoded in eqs.~\eqref{eq:ODE_1}--\eqref{eq:ODE_4}. Figure~\ref{fig:dilution} shows an overview of the evolution of the energy densities $\rho_\text{med}$ and $\rho_\text{rad}$, the normalized scale factor $\bar{a}$, the temperature of the SM bath $T_\text{SM}$, and the amount of injected entropy $S_\text{SM} / S_\text{SM}^\text{cd}$ into the SM bath as functions of the dimensionless time parameter $\theta = t/\tau$. The five physical input parameters are $T_\text{SM}^\text{cd} = 1 \, \text{TeV}$, $m_\text{med} = 100 \, \text{GeV}$, $\tau = 0.1 \, \text{s}$, $\xi_\text{cd} = 1$, and $\alpha_{32} = 0.01$.

We can identify six distinct stages in the evolution.

\textbf{Phase I: Relativistic mediator.} At the initial temperature $T_\text{SM}^\text{cd} = 1 \, \text{TeV}$ the mediators are still relativistic and the universe is dominated by SM radiation such that the scale factor and temperature obey the well-known relations of $a \propto \sqrt{t}$ and $T_\text{SM} \propto 1 / a$. Since the assumed mediator mass is only slightly smaller than the initial dark sector temperature, this phase only lasts for a short period of time.

\textbf{Phase II: Cannibalism.} Once the mediators become non-relativistic, cannibalism processes become relevant and lead to a $\rho_\text{med} \, \bar{a}^3  \propto 1 / \ln \bar{a}$ behaviour. As discussed in section~\ref{subsec:mediatorevo}, this scaling implies a more efficient depletion of the energy density compared to the case without any cannibalistic effects.

\textbf{Phase III: Non-relativistic mediator.} At some point $3\to2$ processes cease to be efficient while decays are not yet relevant, such that the comoving mediator energy density stays constant. Since the energy density of the SM radiation still scales with $\rho_\text{rad} \propto a^{-4}$, the contribution of the mediators becomes increasingly important for the expansion history of the universe.

\textbf{Phase IV: Early matter domination.} Once the mediator density dominates the universe's energy content, we find ourselves in a phase of early matter domination. This means that the scale factor no longer scales as $a \propto t^{1/2}$, but rather with $a \propto t^{2/3}$. Therefore, we can see that $\bar{a}$ starts to deviate from its initial time evolution (marked in light violet in the upper-right panel).
As a result, the temperature of the SM bath falls off slightly more quickly ($T_\text{SM} \propto t^{-2/3}$) than predicted in $\Lambda$CDM ($T_\text{SM} \propto t^{-1/2}$).
			
\textbf{Phase V: Entropy injection.} As soon as the decay term becomes non-negligible, the injection of entropy into the SM bath becomes relevant, influencing the entropy ratio $S_\text{SM} / S_\text{SM}^\text{cd}$ from $\theta \simeq 2 \cdot 10^{-2}$ onwards. The energy density of the mediator slowly starts to decrease as a result of its decay, reheating the SM bath due to $T_\text{SM} \propto \mathcal{S}^{1/4}$, see eq.~\eqref{eq:TSM}. Note, however, that there is no literal ``reheating'' but rather a decrease of the SM temperature that is less steep than $T_\text{SM} \propto a^{-1}$. As shown in reference \cite{Cirelli2018}, the scaling in this period is $T_\text{SM} \propto a^{-3/8}$. With the decrease of mediator energy density, the radiation energy density increases, such that the universe converges towards a radiation dominated era again.

\textbf{Phase VI: Decay.} The decays of the mediator continue after the end of the early matter domination. At $\theta \sim 1$, the two energy densities considered here are again equal and the temperature has almost reached its $\Lambda$CDM evolution, as described by the curve in light red in the bottom-left panel. This curve was calculated using eq.~\eqref{eq:TSM}, but therein setting $a \propto \sqrt{t}$ (as in radiation domination) and $\mathcal{S} = 1$ (for no entropy injection). Once the decaying mediators have injected most of their entropy into the SM bath, the curve for $S_\text{SM} / S_\text{SM}^\text{cd}$ saturates and the subsequent evolution of the universe follows the usual picture for radiation domination.

To conclude this discussion, we note that not all phases described above are present for all parameter points. For example, if $\alpha_{32}$ is very small, there may never be a cannibalism phase. Conversely, if $\alpha_{32}$ is very large, the universe enters a period of cannibal domination rather than matter domination, which only ends when the mediator decays.

\subsection{Dilution of gravitational waves} \label{subsec:dilution}

The out-of-equilibrium decay of the mediator can result in a considerable injection of entropy and energy into the SM bath, as shown in figure~\ref{fig:dilution}, where the comoving entropy of the SM bath after the dark sector decay is more than two orders of magnitude larger than before. While the standard cosmological evolution will be recovered after the mediators have decayed, there is nevertheless one main consequence of the entropy injection: the dilution of frozen-out abundances.

Here we focus on the dilution effect on the generated stochastic GW background, which can be interpreted as a form of dark radiation~\cite{Ellis:2020nnr}. The relevant quantity to quantify the dilution is the scale factor ratio between bubble nucleation and today, see eq.~\eqref{eq:Omegaredshift}. Using the usual entropy-scale factor relation, we obtain
\begin{align}
\frac{a_\text{n}}{a_0} = \frac{1}{D_\text{SM}^{1/3}} \ba{\frac{g_{\text{eff},s}^{\text{SM},0}}{g_{\text{eff},s}^\text{SM,n}}}^{1/3} \frac{T_\text{SM}^0}{T_\text{SM}^\text{n}}\;,
\end{align}
where we have used that the comoving SM entropy is separately conserved before and after the mediator decay. 
We therefore observe once again that an increase of the SM entropy ($D_\text{SM} > 1$) corresponds to a larger scale factor today. The  effects on the frequency spectrum of the GW are then determined by the relation
\begin{align}
\label{eq:FreqRedshift}
f \to \frac{a_\mathrm{n}}{a_0} \, f = D_\text{SM}^{-1/3} \ba{\frac{g_{\text{eff},s}^{\text{SM},0}}{g_{\text{eff},s}^\text{SM,n}}}^{1/3} \frac{T_\text{SM}^0}{T_\text{SM}^\text{n}} \,  f\;,
\end{align}
corresponding to a frequency shift towards smaller values when dilution effects are present.\footnote{We note that an intermediate period of matter domination also leaves a more direct imprint on the GW spectrum by enhancing modes that enter the horizon during that time~\cite{Hook2020}. 
In the case that the period of matter domination happens well after the phase transition, the spectrum close to the peak, and hence the signal-to-noise ratio calculated below, remains however unaffected. We expect a similar conclusion to hold also in the case of an intermediate period of cannibal domination, which has so far not been considered in the literature.}

Importantly, another redshift contribution affects the amplitude of the GW signal directly, which can be included in the $\mathcal{R}$ factor from eq.~\eqref{eq:fancyR}:
\begin{align}
\mathcal{R} &= \frac{1}{D_\text{SM}^{4/3}} \ba{\frac{g_{\text{eff},s}^{\text{SM},0}}{g_{\text{eff},s}^\text{SM,n}}}^{4/3} \frac{\pi^2 \, g_{\text{eff},\rho}^\text{tot,n}}{90} \frac{\ba{T_\text{SM}^0}^4}{\Mp^2 \, H_0^2}\;.\label{eq:RDilut}
\end{align}
We conclude that dilution effects decrease the amplitude of the GW frequency, as visualized in figure~\ref{fig:overviewplot1} together with the frequency shift.
For the following discussion it will be convenient to introduce 
\begin{align}
D \equiv \frac{g_{\text{eff},s}^\text{SM,n}}{g_{\text{eff},s}^\text{tot,n}} \, D_\text{SM}\;, \label{eq:DefD}
\end{align}
which takes into account the contribution of the dark sector to the total energy density (see figure~\ref{fig:geff}).
Using this definition, eq.~\eqref{eq:RDilut} becomes 
\begin{align}
\mathcal{R}\, h^2 \simeq \frac{2.473 \cdot 10^{-5}}{D^{4/3}} \ba{\frac{g_{\text{eff},s}^{\text{SM},0}}{g_{\text{eff},s}^\text{tot,n}}}^{4/3} \frac{g_{\text{eff},\rho}^\text{tot,n}}{2}\;. \label{eq:Rh2}
\end{align}
The advantage of this expression is that for temperature ratios $\xi \gg 1$ we expect $g_{\text{eff},\rho}^\text{tot,n} \propto \xi^4$ and $g_{\text{eff},s}^\text{tot,n} \propto \xi^3$, see eqs.~\eqref{eq:rhotot} and~\eqref{eq:stot}. Hence, the $\xi$ dependence in the degrees of freedom cancels and all effects are encoded entirely in $D$. Moreover, as shown in Ref.~\cite{Cirelli2018}, the dilution factor $D$ also saturates for large $\xi_\text{cd}$, such that in this limit $\mathcal{R}\, h^2$ becomes independent of the temperature ratio between the two sectors.

\begin{figure}
\centering
\includegraphics[width=.9\linewidth]{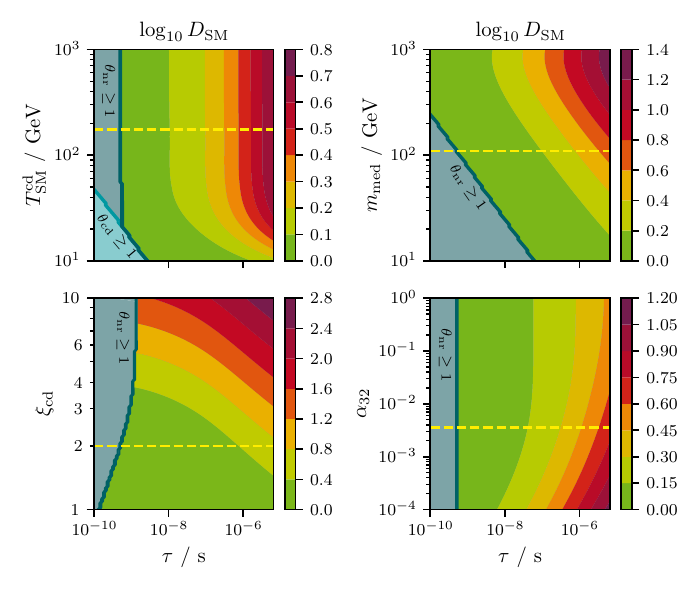}
\caption{
Scan over the five dark sector parameters that determine the dilution factor $D_\text{SM}$. In each panel two parameters are varied explicitly with the other parameters fixed to the benchmark point $m_\text{med} = 110\,\mathrm{GeV}$, $T_\text{SM}^\text{cd} = 175\,\mathrm{GeV}$, $\alpha_{32} = 3.6 \cdot 10^{-3}$ and $\xi_\text{cd} = 2$ (indicated by horizontal dashed lines in each panel). Cyan areas of the plots are excluded because mediator decays occur already during chemical decoupling ($\theta_\text{cd} > 1$), while blue areas correspond to relativistic mediator decays ($\theta_\text{nr} > 1$). Note the change of the colour scale between the different panels.
}
\label{fig:DSMLISA}
\end{figure}

We are now in the position to describe the effect of the mediator decay on the stochastic GW background by investigating the dependence of the dilution factor $D_\text{SM}$ on the input parameters defined at the end of section \ref{subsubsec:derivingODEs}. For a first impression of the impact of the different quantities that specify the dark sector and its decay, we show in figure~\ref{fig:DSMLISA} the results from scans over different planes in the resulting parameter space. More specifically, we show the dependence of $D_\text{SM}$ on the temperature of the SM bath at decoupling $T_\text{SM}^\text{cd}$, the mediator mass $m_\text{med}$, the temperature ratio at chemical decoupling $\xi_\text{cd}$, the effective $3 \rightarrow 2$ coupling $\alpha_{32}$ and the mediator lifetime $\tau$.\footnote{Note that, in contrast to $D$ as introduced in eq.~\eqref{eq:DefD}, $D_\text{SM}$ can be calculated without specifying the degrees of freedom of the entire dark sector, such that results presented in this way are more model-independent.} 
The parameters not varied explicitly in each panel are fixed to $m_\text{med} = 110\,\mathrm{GeV}$, $T_\text{SM}^\text{cd} = 175\,\mathrm{GeV}$, $\alpha_{32} = 3.6 \cdot 10^{-3}$ and $\xi_\text{cd} = 2$.
Note that we exclude parameter regions where the mediator decays are already efficient during chemical decoupling ($\theta_\mathrm{cd} \ge 1$) and where the approximation of non-relativistic mediator decays breaks down ($\theta_\mathrm{nr} \ge 1$).

In the top-left panel we consider the dependence of the dilution factor $D_\text{SM}$ on the chemical decoupling temperature $T_\mathrm{SM}^\mathrm{cd}$ and the mediator lifetime $\tau$. We find that the entropy injection (and hence the dilution) only becomes sizeable for a sufficiently long-lived mediator, such that $\theta_\text{nr} \ll 1$ and there is a substantial period of early matter domination. For $T_\mathrm{DS}^\mathrm{cd} > m_\text{med} = 110 \, \mathrm{GeV}$, corresponding to $T_\mathrm{SM}^\mathrm{cd} > m_\text{med} / \xi_\text{cd} = 55 \, \mathrm{GeV}$, the dependence of $D_\text{SM}$ on the chemical decoupling temperature is very mild and results only from changes in the number of relativistic degrees of freedom. For smaller values of $T_\mathrm{SM}^\mathrm{cd}$, on the other hand, $D_\mathrm{SM}$ decreases rapidly. This decrease is a direct consequence of our assumption that the mediator abundance is given by an equilibrium distribution at chemical decoupling and therefore becomes Boltzmann suppressed at small decoupling temperatures. We note, however, that such a Boltzmann suppression is difficult to achieve in realistic models, and that the examples that we will consider in section~\ref{sec:model} always correspond to $T_\mathrm{DS}^\text{cd} > m_\text{med}$.

In the top-right panel we focus on the dependence of $D_\text{SM}$ on the mediator mass. We find that smaller mediator masses lead to a decrease of the dilution factor. The reason is that lighter mediators experience a longer period of relativistic and cannibalistic evolution, such that the duration of the early matter domination and thus the amount of entropy injection decreases. As before, when $m_\text{med} > T_\mathrm{DS}^\text{cd}$, the mediator is Boltzmann suppressed at decoupling, thus reducing the dilution. 

The bottom-left panel shows the effect of varying the temperature ratio $\xi$ at chemical decoupling. Since an increase in $\xi_\text{cd}$ corresponds to an increase in the energy and entropy stored in the dark sector, it is clear that $D_\text{SM}$ grows with increasing $\xi_\text{cd}$. We note that the dilution factor changes more rapidly with $\tau$ for large lifetimes, corresponding to mediator decays during matter domination, than for smaller lifetimes, corresponding to mediator decays during cannibal domination.

The general effect of the cannibalistic era can be observed in the bottom-right panel, which describes the dependence of $D_\text{SM}$ on $\alpha_{32}$. A large effective $3 \rightarrow 2$ coupling means that the number-changing processes stay efficient for a longer period of time. During such a cannibalistic era the comoving mediator energy density $\rho_\text{med} \, a^3$ decreases, such that the universe enters into the phase of early matter domination at a later point for larger $\alpha_{32}$. We emphasize that this is potentially a large effect: Compared to the case where cannibalism is negligible ($\alpha_{32} = 10^{-4}$) the dilution factor can be suppressed by a factor of a few if $\alpha_{32}$ is large. For large $\alpha_{32}$, the dilution factor becomes essentially independent of $\alpha_{32}$, as the mediator decays before cannibalism ends.

To conclude this section, we remind the reader that large dilution factors correspond to small GW signals in the present universe. We have therefore identified two competing effects: Increasing the temperature ratio $\xi$ increases the stochastic GW background produced during a first-order phase transition (see section~\ref{sec:FOPT}) but also leads to larger dilution factors. However, it should be clear from figure~\ref{fig:DSMLISA} that $\xi$ is not the only relevant parameter. In particular, we expect the mediator lifetime to play a decisive role in determining whether increasing $\xi$ leads to an overall enhancement or suppression of GW signals. We will study this question in a more concrete setting in the following section.

\section{Example: Hot dark Higgs bosons}
\label{sec:model}

We are now equipped with all the necessary tools to investigate a particular dark sector model, determine its phase transitions and the resulting stochastic GW background, calculate the decay of a mediator species and the consequent entropy injection, and finally obtain the present-day GW signal. In section \ref{sec:model_def}, we will provide a description of the model that we study in the rest of this section and identify the five parameters relevant to the entire phenomenological discussion. A study of the effects occurring in the different regions of the available parameter space will be presented in section~\ref{sec:explore_param_space}. Finally, we analyze the expected signal-to-noise ratio in our model for LISA and ET for two benchmark points in section \ref{sec:obs_GW}.

\subsection{Model definition}
\label{sec:model_def}

We consider a simple dark sector model given by a dark photon that arises from a new $U(1)_\text{D}$ gauge group under which a complex Higgs field $\Phi = \ba{\phi + i \, \varphi}/\sqrt{2}$ is charged, while the entire SM field content is neutral. The radial mode $\phi$ is the dark Higgs boson and $\varphi$ is the Goldstone boson contributing to the longitudinal mode of the dark photon after symmetry breaking.  The underlying Lagrangian can be written as~\cite{Breitbach2019}
\begin{align}
\mathcal{L} &\supset \left|D_\mu \, \Phi \right|^2 + \left|D_\mu \, H \right|^2 - \frac{1}{4} \, A_{\mu\nu}^\prime \, A^{\prime \mu \nu} - \frac{\epsilon}{2} \, A_{\mu \nu}^\prime \, B^{\mu \nu} - V(\Phi,H) \;,
\label{eq:modellagrangian}
\end{align}
with the covariant derivative
\begin{align}
D_\mu \, \Phi= \ba{\partial_\mu + i \, g \, A_\mu^\prime} \Phi
\end{align}
and the field-strength tensor $X_{\mu\nu} = \partial_\mu X_\nu - \partial_\nu X_\mu$, where $X$ stands for $A'$ (dark photon) or $B$ (SM hypercharge gauge boson). 

The scalar potential involving the dark Higgs field $\Phi$ and the SM Higgs field $H$ is given by~\cite{Curtin2014}
\begin{align}
V_\text{tree}(\Phi,H) = - \mu^2 \, \Phi^\ast \, \Phi + \lambda \, (\Phi^\ast \, \Phi)^2 - \mu_H^2 \, H^\dagger \, H + \lambda_H \, (H^\dagger \, H)^2 + \lambda_p \, (\Phi^\ast \, \Phi) \, (H^\dagger \, H) \;,
\end{align}
where $\mu^2_i$ and $\lambda_i$ are the different quadratic and quartic couplings. We assume that both $\epsilon$ and $\lambda_p$ are sufficiently small that they do not lead to the thermalisation of the dark sector with the SM thermal bath, which also implies that they play a negligible role for the phase transition. We can therefore treat the dark and visible sectors independently at early times and study the dark Higgs potential in terms of the couplings $\lambda$, $g$ and $\mu$. 
The various mass parameters read~\cite{Breitbach2019}
    \begin{align}
    m^2_{A^\prime}(\phi) &= g^2 \, \phi^2\;,\\
    m^2_\phi &= - \mu^2 + 3 \, \lambda \, \phi^2 \;,\\
    m^2_\varphi(h, \phi) &= - \mu^2 + \lambda \, \phi^2 \;.
    \end{align}
At high temperatures, the effective potential given in eq.~\eqref{eq:effective_potential} has a minimum at $\phi = 0$, such that the dark photon is massless. At zero temperature, on the other hand, the minimum of the potential for $\phi$ lies at $v = \mu / \sqrt{\lambda}$, leading to $m_{A'} = g\ v$ and $m_\phi = \sqrt{2 \, \lambda} \,v$. Furthermore, the coupling relevant for number-changing processes is given by $\alpha_{32} = 9 / (2^{1/3} \, \pi) \, \lambda \approx 2.3 \, \lambda$.

We observe that, for $g > \sqrt{2 \, \lambda}$, the dark Higgs boson is lighter than the dark photon and therefore the lightest particle in the dark sector after the phase transition. As we are going to see, this will be the case for all of the interesting regions of parameter space. The dark Higgs boson therefore takes on the role of the decaying mediator particle discussed in chapter~\ref{sec:dilution}, which is responsible for the energy transfer from the dark sector to the SM. In the following, we will treat the dark Higgs lifetime $\tau$ as an independent parameter that may be determined by $\lambda_p$ or some other unspecified mechanism and assume that the kinetic mixing parameter $\epsilon$ is sufficiently small to neglect dark photon decays into the SM. 
In this set-up the chemical decoupling temperature is approximately given by the temperature when the dark photons become Boltzmann-suppressed: $T_\text{DS}^\text{cd} = m_{A'} > m_\phi$.\footnote{We have checked explicitly that this assumption presents a conservative estimate on the expected signal-to-noise ratio. For $T_\text{DS}^\text{cd} = C \, m_{A^\prime}$ with $0.1 < C < 1$, the chemical decoupling occurs later, such that the dilution factors $D$ will become smaller and the stochastic GW background will be less diluted and thus more easily observable.}

Another important quantity is the temperature ratio $\xi_\mathrm{n}$, which is fixed at the time of nucleation and evolve to later times by tracking the relevant degrees of freedom, see eq.~\eqref{eq:selfconsistentxi}. While we remain agnostic about the precise process that leads to different temperatures of the dark and visible sector, simple possibilities would be a difference of the number of degrees of freedom much earlier in the universe and/or the decay of a heavy particle species that heats the dark sector. In total, our set-up is therefore characterised by five parameters: the dark Higgs quartic coupling $\lambda$, the $\text{U}(1)_\text{D}$ gauge coupling $g$, the vev $v$, the dark Higgs lifetime $\tau$ and the temperature ratio $\xi_\text{n}$ at the nucleation time of the first-order phase transition.
    
In practice, we first compute the details of the first-order phase transition, i.\,e.\ we determine the nucleation temperature $T_\text{DS}^\text{n}$, the inverse time scale $\beta/H$ and the transition strength $\alpha$. In a second step, the dilution factor $D_\text{SM}$ is computed as described in the previous chapter with the initial temperature for the evolution given by the time of decoupling of dark photons. As a final step, we then calculate $D$ according to eq.~\eqref{eq:DefD}. Having determined all these quantities, we can then compute the relevant redshift factors given in eqs.~\eqref{eq:FreqRedshift} and~\eqref{eq:Rh2} which enter the final GW spectrum in eq.~\eqref{eq:Omegaredshift}, linking the entire GW evolution from the time of bubble collisions up until the present day. These calculations are performed with a modified version of \textsc{CosmoTransitions}~\cite{Wainwright2011} described in more detail in appendix~\ref{app:updates}. This code is publicly available as \textsc{TransitionListener} at \url{https://github.com/tasicarl/TransitionListener}.

We emphasize that a subtletly arises in this computation. For some parameter points the nucleation temperature is found to be smaller than the dark photon mass in the broken phase. In such a case the number of relativistic degrees of freedom in the dark sector changes discontinuously from four before the phase transition to one after the phase transition.\footnote{We have checked explicity that the dark Higgs boson is always to good approximation relativistic immediately after the phase transition.} Whenever this happens, we set the temperature of chemical decoupling $T_\text{DS}^\mathrm{cd}$ equal to the nucleation temperature $T_\text{DS}^\mathrm{n}$ (rather than to $m_{A^\prime}$) and assume four degrees of freedom for the evaluation of the nucleation criterion in eq.~\eqref{eq:nucleation}. This approximation accounts for our ignorance of out-of-equilibrium effects at the bubble wall. Indeed, the naively expected particle abundances may be modified by two additional processes known as ``bubble filtering'' \cite{Baker2019,Chway2019} and ``bubble expansion production'' of heavy particles \cite{Azatov2020,Azatov2021}.

\subsection{Exploration of model parameter space}
\label{sec:explore_param_space}
	
In order to understand the parameter dependences of our set-up, it is important to realize that three out of the five aforementioned parameters determine the requirement for bubble nucleation: $g$, $\lambda$ and $v$, with the first two having the strongest influence. In the effective potential the barrier height separating the true and false vacua is set by the gauge coupling $g$, the quartic coupling $\lambda$ determines the depth of the tree-level minimum and, finally, the vev $v$ sets the overall temperature scale for the phase transition. Moreover, the value of $v$ enters the calculations through the number of effective degrees of freedom at nucleation (see below for a more detailed discussion). 

\begin{figure}
	\centering
	\includegraphics[width=.9\linewidth]{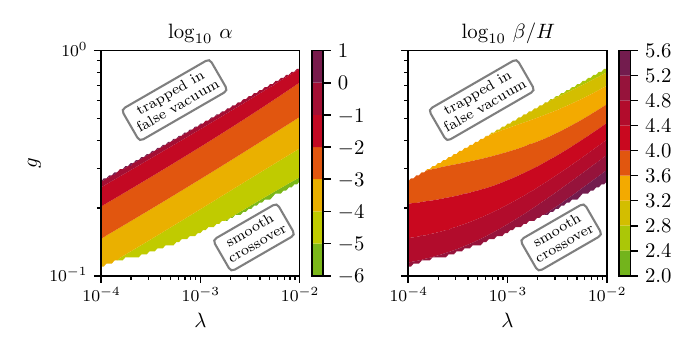}
	\caption{
	Transition strength $\alpha$ and inverse time scale $\beta/H$ in dependence of the $\text{U}(1)_\text{D}$ gauge coupling $g$ and the dark Higgs quartic coupling $\lambda$. The upper boundary of the coloured band corresponds to strong and slow phase transitions. In the white area above, the potential barrier is too high to be overcome, such that bubbles cannot nucleate. In the white area below, a smooth crossover occurs in which no bubbles are emitted either. The tree-level vev was fixed to $v = 2\, \text{TeV}$ and the temperature ratio was set to $\xi_\text{n} = 1$ to generate this figure.}
	\label{fig:LISAlambdag}
\end{figure}

The dependence of the transition parameters $\alpha$ and $\beta/H$ is visualized in figure~\ref{fig:LISAlambdag} (see also Ref.~\cite{Breitbach2019}). As expected, we find that increasing $g$ corresponds to stronger phase transitions, until eventually $g$ is so large that the tunneling rate to the new vacuum receives a great suppression and the universe remains trapped inside of the false vacuum \cite{Biekotter:2021ysx}. For very small values of $g$, conversely, the barrier becomes so small that the phase transition happens smoothly. In figure~\ref{fig:LISAlambdag} the tree-level vev was fixed to $v = 2\, \text{TeV}$, but very similar results would be obtained for somewhat different values. The temperature ratio was set to $\xi_\text{n} = 1$.
	
For the purpose of the following discussion, we are mainly interested in strong phase transitions with large $\alpha$, which also corresponds to an overall slow process with small $\beta/H$. Based on the insight from figure~\ref{fig:LISAlambdag}, the benchmark point $\lambda = 1.5 \cdot 10^{-3}$ and $g=0.5$ fulfills this criterion and will be the focus of the subsequent discussion. We point out, however, that any other point along the upper border of the coloured band would give rise to a qualitatively similar phenomenological discussion presented in the following.

\begin{figure}[t]
\centering
\includegraphics[width=.9\linewidth]{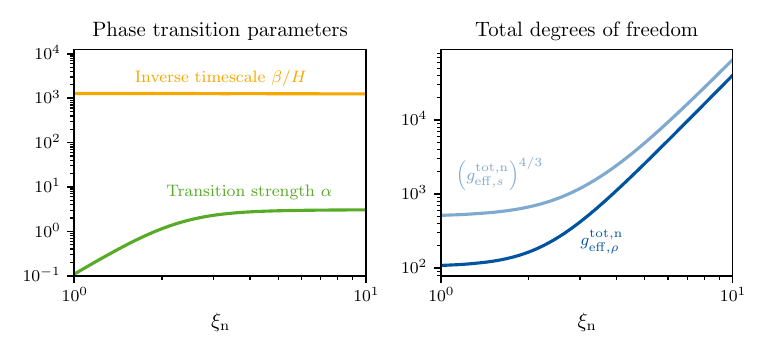}
\caption{
Dependence of the phase transition parameters (left) and the degrees of freedom (right) on the temperature ratio at bubble nucleation. While $\beta$ is independent of $\xi_n$, the transition strength $\alpha$ grows with increasing energy density before saturating.
}
\label{fig:alphabetageff}
\end{figure}

In order to consider the influence of the temperature ratio on the phase transition parameters, we show the dependence of $\alpha$ and $\beta/H$ on $\xi_n$ in figure~\ref{fig:alphabetageff} (left) for $v = 2 \, \mathrm{TeV}$. While $\beta/H$ is insensitive to $\xi_n$~\cite{Breitbach2019}, we observe that even a mild increase from $\xi_n = 1 $ to $\xi_n = 2$ boosts $\alpha$ and therefore the GW spectrum by more than an order of magnitude~-- see the discussion below eq.~\eqref{eq:alpha}. A further increase in the temperature ratio (for fixed $T_\text{DS}^\text{n}$) does not have a large influence on the GW spectrum, because the dark sector begins to dominate the total energy density. Consequently $\rho_\text{tot}^\text{n}$ and therefore $\alpha$ become independent of $\xi_\mathrm{n}$. This feature is further illustrated in the right panel of figure~\ref{fig:alphabetageff}, which shows that the energetic effective degrees of freedom $g_{\text{eff},\rho}^\text{tot,n}$ grow with $\xi_n^4$ for sufficiently large temperature ratios such that the total energy density becomes independent of the SM temperature, see eq.~\eqref{eq:rhotot}. In a similar manner, $\left(g_{\text{eff},s}^\text{tot,n}\right)^{4/3}$ grows with $\xi_n^4$ for large enough $\xi_n$ as can be deduced from eq.~\eqref{eq:stot}, in line with our reasoning below eq.~\eqref{eq:Rh2}.

\begin{figure}[t]
	\centering
	\includegraphics[width=\linewidth]{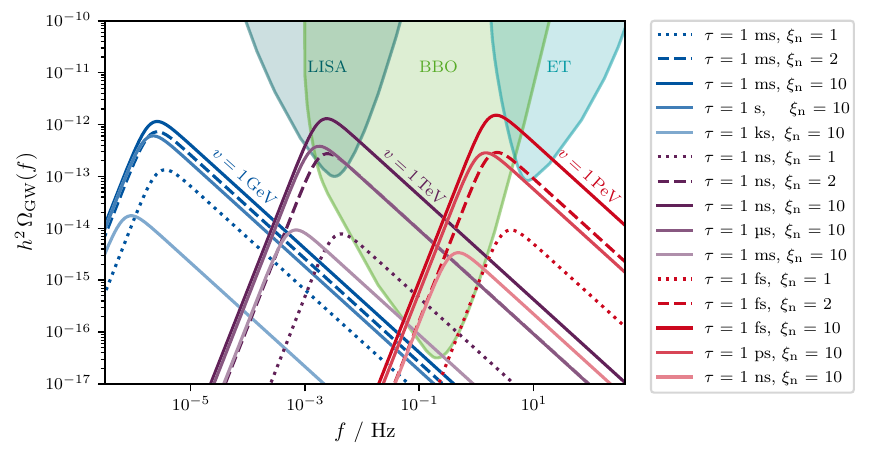}
	\caption{
	An overview plot for the different possible GW spectra that can be provided by our model for a strong first-order phase transition ($\lambda = 1.5 \cdot 10^{-3}$, $g = 0.5$), compared to the expected power-law integrated sensitivity curves for LISA, ET and BBO \cite{Breitbach2019} (see also Ref.~\cite{Alanne:2019bsm,Schmitz:2020syl}). The plot shows the resulting spectra of the phase transition of a dark Higgs acquiring its tree-level vev $v = 1 \, \text{GeV}$ (blue), $v = 1 \, \text{TeV}$ (purple), or $v = 1 \, \text{TeV}$ (red). Dotted lines refer to the case when $\xi_\text{n} = 1$, whereas dashed and solid lines indicate $\xi_\text{n} = 2$ and $\xi_\text{n} = 10$, respectively. The dependence of the spectrum on the dark Higgs lifetime $\tau$ is indicated by lighter colors. The main result is that increasing the temperature ratio $\xi_\text{n}$ leads to a strong enhancement of the signal strength when the dark Higgs decays sufficiently fast. The tree-level vev shifts
	the signal to different frequencies and has a mild impact on the signal strength for $v \le 100 \, \text{GeV}$.}
	\label{fig:overviewplot2}
\end{figure}

In figure~\ref{fig:overviewplot2} we explore the effects of varying the vev $v$, the lifetime $\tau$ and the temperature ratio $\xi_\mathrm{n}$ on the GW spectra. For comparison we show the expected power-law integrated sensitivity curves for LISA, ET and BBO \cite{Breitbach2019}. As expected, the main effect of varying the vev (indicated by the different colours) is to change the nucleation temperature and hence the peak frequency. A change in the temperature ratio $\xi_\mathrm{n}$ for fixed $v$ amounts to altering the GW spectra as visualized by the dotted, dashed, and (dark) solid lines. As already observed in figure~\ref{fig:alphabetageff}, increasing the temperature ratio by a factor of two from $\xi_\mathrm{n} = 1$ to $\xi_\mathrm{n} = 2$ is already sufficient to boost the GW spectrum significantly by one to two orders of magnitude, much to the benefit of experimental prospects.

The influence of the lifetime $\tau$ and hence the dilution effects can be observed in figure~\ref{fig:overviewplot2} by comparing the solid lines of different shading, with dilution effects increasing from darker to lighter shading for fixed $v$ and $\xi_\mathrm{n} = 10$.  As discussed above, increasing the lifetime of the dark Higgs boson corresponds to a larger entropy injection, diluting the signal towards both smaller signal strengths and frequencies, which eventually leads to a reduced experimental sensitivity in these scenarios. Note that we consider different lifetimes of the dark Higgs boson for the different vevs. The reason is that the quantity that sets the relevant scale for $\tau$ in the calculation of $D$ is the Hubble parameter, which scales with $T_\text{DS}^2$. Hence,  to obtain a comparable value for the signal strength and the dilution factor, smaller values of $\tau$ are needed for larger values of $v$, corresponding to larger nucleation temperatures. For the darkest (uppermost) curves, we have chosen $\tau$ such that the dilution factor is of order unity for all three cases.

Let us finally have a closer look at the dependence of the GW spectra on the vev. We observe that for $v = 1\,\mathrm{GeV}$ ($\xi_n=1$) a peak signal strength of $h^2 \,\Omega_\text{GW}\approx 10^{-13}$ is obtained, whereas the corresponding signals for $v = 1\,\mathrm{TeV}$ and $v = 1\,\mathrm{PeV}$ lie at significantly smaller values. The underlying reason for this is that larger values of $v$ correspond to more relativistic degrees of freedom and hence a larger energy density in the SM at nucleation, leading to smaller $\alpha$. However, eventually the degrees of freedom reach their maximal value in the SM at about $v \simeq 100\,\mathrm{GeV}$, beyond which $\alpha$ does not change considerably with the vev anymore. For our example, we have $\alpha = 0.51$ for  $v = 1\,\mathrm{GeV}$ and $\alpha = 0.11$ for  $v = 1\,\mathrm{TeV},\,1\,\mathrm{PeV}$ when considering an equally hot dark and visible sector ($\xi_n=1$).

Given the scaling $h^2 \, \Omega_\text{GW} \propto \alpha^2/(1+\alpha)^2$~\cite{Huber:2008hg,Breitbach2019} we are also able to deduce that the enhancement of the GW signal with increasing temperature ratio is stronger the weaker the initial spectrum, i.\,e.\ if the initial value of $\alpha$ is rather small. Therefore, the effect of increasing $\xi_\mathrm{n}$ is larger for $v=1\,\mathrm{TeV},\,1\,\mathrm{PeV}$ than for $v=1\,\mathrm{GeV}$. This consideration also explains the fact that all spectra converge towards a comparable peak signal strength with large $\alpha$ for $\xi_n=10$ despite different underlying values for the vev.

An important conclusion that can be drawn from figure~\ref{fig:overviewplot2} is that TeV- and PeV-scale vevs are most favorable for LISA and ET, respectively. We emphasize, however, that far-future GW interferometers such as BBO~\cite{Crowder:2005nr}, which would operate in an intermediate frequency range with very high sensitivity, would also be sensitive to these benchmark scenarios.  For the following section, we will take a closer look at the two benchmark values $v_1=2\,\mathrm{TeV}$ and $v_2=10\,\mathrm{PeV}$ for $\lambda = 1.5 \cdot 10^{-3}$ and $g=0.5$. In these cases, the nucleation temperatures lie at $T^\text{n}_{\text{DS},1} = 175\,\mathrm{GeV}$ and \mbox{$T^\text{n}_{\text{DS},2} = 851\,\mathrm{TeV}$}, respectively, while the dark photon mass is given by $m_{A',1} = 1\,\mathrm{TeV}$ and $ m_{A',2} = 5\,\mathrm{PeV}$. Consequently, we encounter the scenario mentioned before: the dark photon decouples on a very short timescale once the universe enters the new phase. As discussed above, we therefore set $T^\text{cd}_\text{DS} = T^\text{n}_\text{DS}$ and identify the temperature ratio at nucleation $\xi_n$ with the one at chemical decoupling $\xi_\text{cd}$. 

\subsection{Sensitivity for gravitational waves}
\label{sec:obs_GW}

In this section, we discuss the prospects for LISA and ET for the two benchmark points described above. While figure~\ref{fig:overviewplot2} allows to assess the observability of a GW signal in a qualitative manner, an analysis based on signal-to-noise ratios is in fact better suited to quantify the experimental sensitivities~\cite{Breitbach2019}. In this discussion, we aim to study the two main competing effects in particular: the enhancement of GWs at production through a large temperature ratio $\xi_\mathrm{n}$ and the subsequent dilution of the spectrum determined primarily by the dark Higgs lifetime $\tau$. In the following, we therefore analyse the signal-to-noise ratios in terms of these two quantities for fixed values of the other three parameters.

For the calculation of signal-to-noise ratios, we employ the auto-correlated optimal-filter measure
\begin{align}
	\text{SNR} = \sqrt{t_\text{obs} \, \int_{f_\text{min}}^{f_\text{max}} \, \text{d}f \; \bb{\frac{h^2 \, \Omega_\text{GW}(f)}{h^2 \, \Omega_\text{eff}(f)}}^2} \; ,
\end{align}
as derived in Ref.~\cite{Breitbach2019}. The observed signal strength $\Omega_\text{GW}$ (see eq.~\eqref{eq:Omegaredshift}) is normalized to the effective noise energy density spectrum $\Omega_\text{eff}$, integrated within the frequency band $\bb{f_\text{min}, f_\text{max}}$ in which the detector is sensitive, and weighted with the duration of observation $t_\text{obs}$. The noise spectrum not only encompasses instrumental noises but also noise from unresolved galactic binaries. We will refer to the signals as being detectable by LISA and the Einstein Telescope for signal-to-noise ratios exceeding the respective threshold SNR values of 10 and 5. A detailed overview of the calculation of signal-to-noise ratios as well as the used noise curves can be obtained from Ref.~\cite{Breitbach2019}.

\begin{figure}
	\centering
	\includegraphics[width=.82\linewidth]{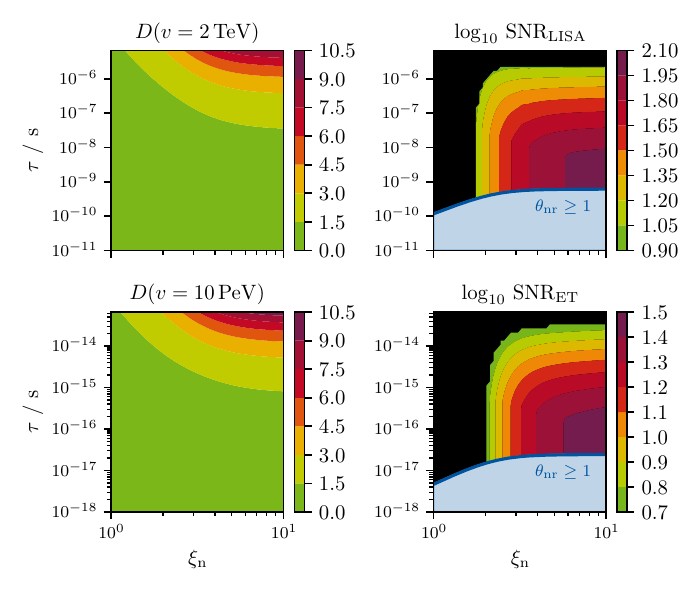}
	\caption{
		Dependence of the dilution factor (left) and the expected signal-to-noise ratios in future GW observatories (right) as a function of the temperature ratio at bubble nucleation and the dark Higgs lifetime. In the top row we consider a Higgs vev $v = 2 \, \mathrm{TeV}$, corresponding to potentially observable signals in LISA, while in the bottom row we choose $v = 10 \, \mathrm{PeV}$ and consider ET. In both cases we find that large temperature ratios significantly enhance the predicted strength of the GW signal, provided the dark Higgs lifetime is short enough to avoid significant dilution.
	}
	\label{fig:SNRsLISAET}
\end{figure}

In figure~\ref{fig:SNRsLISAET} we visualize our main results for the LISA and ET benchmark points. In the left column, the dependence of the dilution factor $D$ on $\xi_\mathrm{n}$ and $\tau$ is shown, while the right-hand side focuses on the signal-to-noise ratios in the same parameter plane. For both LISA and ET we obtain qualitatively similar situations. For equally hot dark and visible sectors ($\xi_n=1$), the GWs from the first-order phase transition are not observable irrespective of what the dark Higgs lifetime is. Only when increasing the temperature ratio to $\xi_n \gtrsim 2$, amounting to more energy stored in the GWs, we reach parameter regions for which both experiments become sensitive, thus validating our initial motivation that hot dark sectors greatly increase the observability of GWs. However, we also observe that for larger lifetimes, eventually, the signals get weaker again in spite of large $\xi_n$. This is the case because dilution effects become increasingly important the larger the dark Higgs lifetime is, redshifting the GW spectrum and decreasing the experimental reach for such scenarios. The sensitivity loss for larger $\tau$ is also independent of $\xi_n$ for large enough temperature ratios since both $\alpha$ and $D$ saturate eventually, leading to a nearly horizontal turnover of the LISA and ET sensitivity. This interplay between enhancing and diluting the GWs ultimately results in a rectangular shape of the relevant signal-to-noise ratio region in the right column.

We emphasize that in the most interesting parameter regions the lifetime of the dark Higgs boson is sufficiently small to not interfere with BBN. However, given our assumption that the dark Higgs boson decays non-relativistically, the lifetime cannot be arbitrarily small. This is indicated by the blue shaded region in the right column, which corresponds to $\theta_\mathrm{nr} \geq 1$ (see section~\ref{sec:dilution}). Based on the left column of figure~\ref{fig:SNRsLISAET}, however, we do not expect dilution effects to become relevant in this parameter region, which would correspond to the case that the universe never enters a period of early matter domination. 

Taking all relevant effects and conditions into account, we are therefore able to find sizeable parameter regions in $\xi_n$ and $\tau$ that predict sufficiently high signal-to-noise ratios to offer attractive prospects for future GW observatories. For LISA the signal-to-noise ratio can be as large as $\mathcal{O}(100)$, while the corresponding values for ET are somewhat smaller due to its steeper sensitivity loss for small frequencies. This finding demonstrates clearly that GW signatures, which may seem out of reach, can in fact be enhanced as soon as the dark sector is hot, enabling one to distinguish GWs emerging in the early universe from noise affecting the experimental measurement. Further progress in testing the GW scenarios discussed here can be expected from BBO, which will also be able to cover vevs within the TeV-PeV range and also be sensitive to scenarios with stronger dilution effects (see figure~\ref{fig:overviewplot2}).

To conclude this discussion, we emphasize once more that the enhancement and dilution effects considered in this work apply to a wide range of GW signals from dark sector phase transitions. In particular, our findings can be directly applied to different values of $\lambda$ and $g$, which would lead to smaller $\alpha$ and larger $\beta/H$, corresponding to overall weaker GW signals. In this case a larger vev would be required to achieve a similar peak frequency and to counterbalance the change in the nucleation temperature. At the same time, the effect from increasing $\xi_\mathrm{n}$ would be even larger, given that the saturation of $h^2 \, \Omega_\text{GW} \propto \alpha^2/(1+\alpha)^2$ would be delayed. Hence, for sufficiently large temperature ratios $\xi_n$ even such comparably weak GW signals may be rendered observable. 

\section{Conclusions}
\label{sec:conclusions}

In this work we have considered the exciting prospect that future gravitational wave (GW) observatories will be able to measure the stochastic GW background from first-order phase transitions. Such first-order phase transitions arise frequently in extensions of the Standard Model (SM) that feature a dark sector, i.\,e.\ a collection of new states that interact with each other but only very weakly with the SM. A crucial property of such a dark sector is that it may have a temperature different from the temperature of the SM thermal bath. The hotter the dark sector, the larger its contribution to the total energy density and therefore the stronger the GW signals that can be produced. In this work we have focused on the case that the dark sector has a larger temperature than the visible sector, which can significantly boost GW signals that would otherwise be unobservable.

Such a set-up however faces a great challenge: In order to recover the standard cosmological evolution at low temperatures, the entropy stored in the dark sector needs to be transferred to the visible sector. To achieve this goal we have considered the case that the lightest dark sector particle (called the mediator) is unstable against decays into SM particles. These decays must be sufficiently slow that they do not bring the two sectors into thermal equilibrium at early times. On the other hand, if these decays happen too late, the universe enters a phase of early matter domination and the eventual transfer of entropy to the visible sector leads to a strong dilution of the GW signal.

We have explored the dependence of this dilution effect on the properties of the mediator (specifically its mass, lifetime, decoupling temperature and self-interactions) and the dark sector temperature. We have extended previous works on the topic by considering a period of cannibalism, during which the energy in the dark sector decreases through number-changing processes. Such cannibalism can significantly reduce the resulting dilution factors and thereby extend the range of mediator lifetimes, for which GW signals may be observable. 

To apply our findings to a realistic scenario, we have considered a dark sector describing the spontaneous breaking of a new $U(1)_\mathrm{D}$ gauge symmetry with a dark Higgs field and a dark photon as field content. In the parameter regions where a strong first-order phase transition is predicted, the lightest dark sector particle is the dark Higgs boson, while the dark photon obtains a mass larger than the dark sector temperature at bubble nucleation. If the dark sector temperature is equal to the SM temperature, the predicted GW signals are below the sensitivity of next-generation GW observatories such as LISA or the Einstein Telescope. However, temperature ratios of order 2 are sufficient to boost the GW signal above the expected level of noise. The subsequent dilution of the signal due to entropy injection remains small provided the dark Higgs bosons decay sufficiently quickly after the phase transition. We find that the interesting regions of parameter space span several orders of magnitude in the dark Higgs lifetime. Further regions of parameter space are expected to open up when extending our analysis to the case that the dark Higgs bosons decay before they become non-relativistic.  The code used to obtain our results is publicly available as \textsc{TransitionListener} at \url{https://github.com/tasicarl/TransitionListener}.

We emphasize that our calculation of the actual GW signals and the resulting signal-to-noise ratios is rather simplified and could be improved in a number of ways, for example by considering the percolation temperature instead of the nucleation temperature or by including effects such as bubble filtering or the production of heavy particles during bubble expansion. Nevertheless, our central findings are independent of these approximations: The enhancements that we find for large dark sector temperatures and the dilution factors that we calculate are independent of the details of the phase transition and can equally be applied to more refined calculations. Our largely model-independent treatment facilitates the transfer of our results to different settings.

For the specific dark sector model that we consider it will be interesting to further explore the connection to dark matter. To do so, it will be essential to specify in detail the couplings of the dark sector to the SM. For example, if kinetic mixing between the dark photon and hypercharge gauge bosons is absent, the dark photon itself could be stable and a viable dark matter candidate. Alternatively, the dark sector could contain additional fermions that freeze out before the phase transition. Such a set-up might furthermore provide an explanation for the difference in temperature between the dark and the visible sector that we have assumed from the beginning. Observatories such as LISA and the Einstein Telescope may therefore have a unique possibility to combine both dark sector and GW physics, with the experimental sensitivity enhanced greatly when hot dark sectors are involved.

\acknowledgments

We thank Moritz Breitbach, Jose Eliel Camargo-Molina, Saniya Heeba, Thomas Konstandin, Julien Lesgourgues and Ville Vaskonen for discussions and Yann Gouttenoire and Filippo Sala for helpful comments on the manuscript. This  work  is  funded  by  the  Deutsche Forschungsgemeinschaft (DFG) through the Emmy Noether Grant No.\ KA 4662/1-1.

\begin{appendix}

\section{Derivation of the equations governing entropy injection} \label{app:derivingODEs}

We start the derivation by noting that the entropy of the SM in a comoving volume $a^3$ is given by $S_\text{SM}(t) = \frac{2 \pi^2}{45} \, \gsSM(t) \, T_\text{SM}^3(t) \, a^3(t)$, which gives
\begin{align}
T_\text{SM}(t) = \ba{\frac{45}{2 \pi^2}}^{1/3}\ba{ \frac{ S_\text{SM}(t)}{\gsSM(t)}}^{1/3} \frac{1}{a(t)} \; . \label{eq:SMtemp}
\end{align}
The energy density of the SM radiation can therefore be expressed as
\begin{align}
\rho_\text{rad}(t) &= \frac{\pi^2}{30} \, \grhoSM(t)  \, T_\text{SM}^4(t) = \frac{3}{4} \ba{\frac{45}{2 \pi^2}}^{1/3} \ba{\frac{S_\text{SM}(t)}{\gsSM(t)}}^{4/3} \frac{\grhoSM(t)}{a^4(t)} \; , \label{eq:SMrad}
\end{align}
allowing us to relate the energy density at chemical decoupling to later times by
\begin{align}
\frac{\rho_\text{rad}(t)}{\rho_\text{rad}^\text{cd}} = \frac{\grhoSM(t)}{\grhoSMcd} \ba{\frac{\gsSMcd}{\gsSM(t)}}^{4/3} \ba{\frac{S_\text{SM}(t)}{S_\text{SM}^\text{cd}}}^{4/3} \bar{a}^{-4} \; . \label{eq:rhoradratio}
\end{align}

Switching to the dimensionless time parameter $\theta = t/\tau$, the first Friedmann equation reads $H^2 = \ba{\frac{\bar{a}^\prime}{\tau \, \bar{a}}}^2 = \frac{\rho_\text{tot}}{3 \, \Mp^2}$ with $\rho_\text{tot} = \rho_\text{med} + \rho_\text{mat} + \rho_\text{rad}$. Here, $\rho_\text{mat}$ describes the influence of any decoupled non-relativistic species, such as a frozen-out dark matter component, which therefore scales as $\rho_\text{mat} = \rho_\text{mat}^\text{cd} \,\bar{a}^{-3}$. The Friedmann equation thus reads
\begin{align}
\bar{a}^\prime(\theta) &= \tau \, \bar{a}(\theta) \sqrt{\frac{\rho_\text{med}(\theta) + \rho_\text{mat}(\theta) + \rho_\text{rad}(\theta)}{3 \,\Mp^2}}\\
&=  \frac{\bar{a}(\theta)}{\theta_\text{H}} \sqrt{\frac{\rho_\text{med}(\theta)}{\rho_\text{med}^\text{cd}} + r_\text{rad}^\text{cd} \, \frac{\rho_\text{mat}(\theta)}{\rho_\text{mat}^\text{cd}} + r_\text{mat}^\text{cd} \, \frac{\rho_\text{rad}(\theta)}{\rho_\text{rad}^\text{cd}}}\; ,
\end{align}
where we have introduced $r_\text{mat}^\text{cd} \equiv \rho_\text{mat}^\text{cd} / \rho_\text{med}^\text{cd}$, $r_\text{rad}^\text{cd} \equiv \rho_\text{rad}^\text{cd} / \rho_\text{med}^\text{cd}$ and $\theta_\text{H} \equiv \sqrt{3 \, \Mp^2 / (\tau^2 \, \rho_\text{med}^\text{cd})}$. Employing eq.~\eqref{eq:rhoradratio} and defining $r = \rho_\text{med} / \rho_\text{med}^\text{cd}$, the first Friedmann equation can be expressed as
\begin{align}
\bar{a}^\prime &= \frac{\bar{a}}{\theta_\text{H}} \sqrt{r + \frac{r_\text{mat}^\text{cd}}{\bar{a}^3} + \frac{r_\text{rad}^\text{cd}}{\bar{a}^4} \frac{\grhoSM}{\grhoSMcd} \ba{\frac{\gsSMcd}{\gsSM}}^{4/3} \ba{\frac{S_\text{SM}}{S_\text{SM}^\text{cd}}}^{4/3}}, \label{eq:Friedmanndimless}
\end{align}
where all of $\bar{a}$, $r$, $\grhoSM$, $\gsSM$ and $S_\text{SM}$ depend implicitly on $\theta$. The last term in eq.~\eqref{eq:Friedmanndimless} describes the amount of entropy that has been injected into the SM bath since $T_\text{SM} = T_\text{SM}^\text{cd}$, thus increasing the radiation energy density therein. 

In order to derive a second differential equation quantifying the entropy injection, we point out the relation
\begin{align}
\td{}{\theta} \bb{\ba{\frac{S_\text{SM}(\theta)}{S_\text{SM}^\text{cd}}}^{4/3}} = \frac{4}{3} \ba{\frac{S_\text{SM}(\theta)}{S_\text{SM}^\text{cd}}}^{1/3} \frac{S_\text{SM}^\prime (\theta)}{S_\text{SM}^\text{cd}} \; . \label{eq:derivativeinjection}
\end{align}
As the SM and dark sector heat fulfill $Q_\text{SM}^\prime = - Q_\text{DS}^\prime$, see eq.~\eqref{eq:vol_heat} and the surrounding text, we can deduce that
\begin{align}
S_\text{SM}^\prime(\theta) &= \frac{Q_\text{SM}^\prime(\theta)}{T_\text{SM}(\theta)} = - \frac{Q_\text{DS}^\prime(\theta)}{T_\text{SM}(\theta)} = -\frac{q_\text{DS}^\prime(\theta) \ a^3(\theta)}{T_\text{SM}(\theta)} \nonumber\\
&= - \ba{\frac{2\,\pi^2}{45}}^{1/3} \ba{\frac{\gsSM(\theta)}{S_\text{SM}(\theta)}}^{1/3} q_\text{DS}^\prime (\theta)  \, a^4(\theta) \; ,
\end{align}
where we have used $Q_\text{DS}^\prime = q_\text{DS}^\prime \, a^3$ along with eq.~\eqref{eq:SMtemp}. Inserting this expression into eq.~\eqref{eq:derivativeinjection} and relating the SM entropy at chemical decoupling to the energy density via eq.~\eqref{eq:SMrad}, we find
\begin{align}
\td{}{\theta} \bb{\ba{\frac{S_\text{SM}}{S_\text{SM}^\text{cd}}}^{4/3}} &= - \frac{4}{3}\ba{\frac{2\, \pi^2}{45}}^{1/3} \bb{\frac{\gsSM}{\ba{S_\text{SM}^\text{cd}}^4}}^{1/3} q_\text{DS}^\prime\,a^4 \\
&= - \bb{\frac{\gsSM(\theta)}{\gsSMcd}}^{1/3} \frac{\grhoSMcd}{\gsSMcd} \frac{q_\text{DS}^\prime(\theta)}{\rho_\text{rad}^\text{cd}} \bar{a}\,^4(\theta) \;. \label{eq:Senquation}
\end{align}

Together with the Friedmann eq.~\eqref{eq:Friedmanndimless}, we have derived a set of coupled differential equations that describe the evolution of the scale factor and the entropy in the SM bath. The system is, however, still under-determined since the time evolution of the degrees of freedom is not trivial. We therefore introduce the functions
\begin{align}
\gamma(\theta) \equiv \frac{\grhoSM (\theta)}{\gsSM (\theta)} \; ,&&  
\mathcal{G}(\theta) \equiv \frac{\gsSM(\theta)}{\gsSMcd}\; ,&&
\mathcal{S}(\theta) \equiv \ba{\frac{S_\text{SM}(\theta)}{S_\text{SM}^\text{cd}}}^{4/3}\;,
\end{align}
and note that the time evolution of the first two functions is described by
\begin{subequations}
\begin{align}
\gamma^\prime(\theta) &= \td{}{T_\text{SM}} \left.\bb{\frac{\grhoSM(T_\text{SM})}{\gsSM(T_\text{SM})}}\right|_{T_\text{SM}(\theta)} T_\text{SM}^\prime \equiv \hat{\gamma}(\theta) \, T_\text{SM}^\prime(\theta) \; , \label{eq:gammaprime}\\
\mathcal{G}^\prime(\theta) &= \td{}{T_\text{SM}} \left.\bb{\frac{\gsSM(T_\text{SM})}{\gsSMcd}}\right|_{T_\text{SM}(\theta)} T_\text{SM}^\prime \equiv \hat{\mathcal{G}}(\theta) \, T_\text{SM}^\prime(\theta) \; .	\label{eq:Gprime}
\end{align}
\end{subequations}
Both $\hat{\gamma}$ and $\hat{\mathcal{G}}$ can thus be calculated from the known temperature evolution of the effective SM degrees of freedom and eq.~\eqref{eq:SMtemp} for a given time $\theta$. 

In order to obtain a closed system of coupled differential equations, we further need to describe $T_\text{SM}^\prime$. Dividing eq.~\eqref{eq:SMtemp} by the corresponding expression evaluated at $\theta = \theta_\text{cd}$ yields
\begin{align}
\frac{T_\text{SM}(\theta)}{T_\text{SM}^\text{cd}} &= \frac{a}{a_\text{cd}} \ba{\frac{S_\text{SM}}{S_\text{SM}^\text{cd}}}^{1/3} \ba{\frac{\gsSMcd}{\gsSM}}^{1/3} = \frac{\bar{a} \, \mathcal{S}^{1/4}}{\mathcal{G}^{1/3}}\\
\Rightarrow \td{}{\theta} \frac{T_\text{SM}(\theta)}{T_\text{SM}^\text{cd}} &= \frac{3\, \mathcal{G} \, \bar{a} \, \mathcal{S}^\prime - 12\, \mathcal{G} \, \bar{a}^\prime \, \mathcal{S} - 4 \,\mathcal{G}^\prime \, \bar{a} \, \mathcal{S}}{12 \,\mathcal{G}^{4/3} \, \mathcal{S}^{3/4} \, \bar{a}^2} \; .
\end{align}
Note that the time evolution of $T_\text{SM}$ itself depends on $\mathcal{G}^\prime$. Inserting the expression just obtained into eq.~\eqref{eq:Gprime} and solving for $\mathcal{G}^\prime$ yields
\begin{align}
\mathcal{G}^\prime(\theta) = - \frac{3}{4} \, \frac{T_\text{SM}^\text{cd} \, \mathcal{G} \, \hat{\mathcal{G}}}{\mathcal{S}^{3/4} \, \bar{a}} \,  \frac{4 \, \mathcal{S} \, \bar{a}^\prime - \mathcal{S}^\prime  \, \bar{a}}{T_\text{SM}^\text{cd} \, \hat{\mathcal{G}} \,  \mathcal{S}^{1/4} + 3 \, \mathcal{G}^{4/3} \, \bar{a}} \; .
\end{align}
Since $\mathcal{G}^\prime$ is now determined, we can also use eq.~\eqref{eq:gammaprime} to describe the time evolution of $\gamma(\theta)$:
\begin{align}
\gamma^\prime(\theta) = \hat{\gamma} \, T_\text{SM}^\text{cd}   \, \frac{3\, \mathcal{G} \, \bar{a} \, \mathcal{S}^\prime - 12\, \mathcal{G} \, \bar{a}^\prime \, \mathcal{S} - 4 \,\mathcal{G}^\prime \, \bar{a} \, \mathcal{S}}{12 \,\mathcal{G}^{4/3} \, \mathcal{S}^{3/4} \, \bar{a}^2} \; .
\end{align}

The final piece needed to obtain a fully determined set of differential equations can be obtained by including a specific time evolution of the mediator species, as it has been discussed in section \ref{subsec:mediatorevo}. Consider that the volume heat rate of our dark sector is given by
\begin{align}
q_\text{DS}^\prime(\theta) = \rho_\text{med}^\prime(\theta) + 3 \, \frac{\bar{a}^\prime(\theta)}{\bar{a}(\theta)} \, \zeta(\theta) \, \rho_\text{med}(\theta) = - \rho_\text{med}(\theta) \; .
\end{align}
The factor $q_\text{DS}^\prime / \rho_\text{rad}^\text{cd}$, which is of relevance for the entropy injection in eq.~\eqref{eq:Senquation}, can therefore be simplified to
\begin{align}
\frac{q_\text{DS}^\prime(\theta)}{\rho_\text{rad}^\text{cd}} = - \frac{\rho_\text{med}(\theta)}{\rho_\text{rad}^\text{cd}} = - \frac{\rho_\text{med}^\text{cd}}{\rho_\text{rad}^\text{cd}} \, r(\theta) =- \frac{r(\theta)}{r_\text{rad}^\text{cd}} \; ,
\end{align}
where
\begin{align}
r^\prime(\theta) = - r(\theta) - 3 \,  \frac{\bar{a}^\prime(\theta)}{\bar{a}(\theta)} \, \zeta(\theta) \, r(\theta)\;.
\end{align}

The time evolution of the scale factor, the SM entropy, the mediator energy density and the effective degrees of freedom in the SM bath can therefore be described by the following set of coupled differential equations:
\begin{align}
\bar{a}^\prime &= \frac{\bar{a}}{\theta_\text{H}} \sqrt{r + \frac{r_\text{mat}^\text{cd}}{\bar{a}^3} + \frac{r_\text{rad}^\text{cd}}{\bar{a}^4} \,  \frac{\gamma}{\gamma_\text{cd}} \, \frac{\mathcal{S}}{\mathcal{G}^{1/3}}} \; , \nonumber\\
\mathcal{S}^\prime &= \frac{r \, \bar{a}^4}{r_\text{rad}^\text{cd}} \, \mathcal{G}^{1/3} \, \gamma_\text{cd} \; , \nonumber \\
r^\prime &= - r - 3 \, \frac{\bar{a}^\prime}{\bar{a}} \, \zeta \, r \; , \label{eq:app_full_ODE_set}\\
\mathcal{G}^\prime &= - \frac{3}{4} \, \frac{T_\text{SM}^\text{cd} \, \mathcal{G} \, \hat{\mathcal{G}}}{\mathcal{S}^{3/4} \, \bar{a}} \, \frac{4 \, \mathcal{S} \, \bar{a}^\prime - \mathcal{S}^\prime  \, \bar{a}}{T_\text{SM}^\text{cd} \, \hat{\mathcal{G}} \,  \mathcal{S}^{1/4} + 3 \, \mathcal{G}^{4/3} \bar{a}} \; , \nonumber\\
\gamma^\prime &= \hat{\gamma} \, T_\text{SM}^\text{cd} \,  \frac{3\, \mathcal{G} \, \bar{a} \, \mathcal{S}^\prime - 12\, \mathcal{G} \, \bar{a}^\prime \, \mathcal{S} - 4 \,\mathcal{G}^\prime \, \bar{a} \, \mathcal{S}}{12 \,\mathcal{G}^{4/3} \, \mathcal{S}^{3/4} \, \bar{a}^2} \; . \nonumber
\end{align}
The equations used in the main text are then obtained by neglecting the contribution from non-relativistic matter ($r_\text{mat}^\text{cd} = 0$) and focusing on temperatures above the MeV-scale, for which $\gamma = 1$.

\section{Modifications of CosmoTransitions}
\label{app:updates}

To perform our analysis of possible phase transitions in dark sectors, we used a customized version of \textsc{CosmoTransitions}~\cite{Wainwright2011}. \textsc{CosmoTransitions} comes with the necessary tools to trace the global and local minima of a given effective potential of one or multiple scalar fields. Moreover, it allows to identify the possible phase transitions between these minima. \textsc{CosmoTransitions} is often used as a benchmark code in the literature \cite{Caprini2019,Camargo-Molina2013}, as it is sufficiently stable and fast. Apart from the identification of first-order phase transitions, also the calculation of bounce actions and bubble profiles is possible with \textsc{CosmoTransitions}.

We first updated the individual modules of the program to work with \textsc{Python} 3 and extended it by an accurate nucleation criterion for first-order phase transitions in dark sectors with a temperature different from the SM bath. Next, we added the code necessary to compute the important phase transition parameters $\alpha$ and $\beta/H$. This requires a model file, in which the effective potential $V_\text{eff}^\mathrm{1-loop}(\phi)$ and the mass spectrum of the dark sector are given, and an additional module to calculate the effective degrees of freedom. Furthermore, we added a module for the calculation of dilution factors $D_\text{SM}$, in which the decay of the dark sector is modeled. Another module for the calculation of stochastic gravitational wave spectra and the signal-to-noise ratios has been added to interpret the observability of the generated signals. This set of modules is controlled by an interface, which itself is executed by a small scan file, which defines the region of parameter space that one wishes to analyze. In addition, there are a few parameters for adjusting the settings, such as the accuracy of scans and the grid for the scan.

\end{appendix}

\bibliographystyle{JHEP_improved}
\bibliography{PTinHotDS.bib}

\end{document}